\documentclass[pdftex,preprint,number]{elsarticle}

\usepackage{hyperref}
\usepackage{algorithm2e}
\usepackage{floatrow}

\usepackage{natbib}

\usepackage{soul}

\usepackage{caption, subcaption, graphicx}
\usepackage{amsmath,bm}
\usepackage{fancyhdr}
\usepackage{graphicx}
\usepackage[margin=2.0cm]{geometry}

\journal{International Journal of Multiphase Flow}

\begin{document}

\begin{frontmatter}
\title{Three-dimensional high speed drop impact \\ onto solid surfaces at arbitrary angles}

\author[mymainaddress]{Radu Cimpeanu\corref{mycorrespondingauthor}}
\cortext[mycorrespondingauthor]{Corresponding author}
\ead[url]{www.imperial.ac.uk/people/radu.cimpeanu11}
\author[mymainaddress]{Demetrios T. Papageorgiou}

\address[mymainaddress]{Department of Mathematics, Imperial College London, SW7 2AZ, London, United Kingdom}

\begin{abstract}

The rich structures arising from the impingement dynamics of water 
drops onto solid substrates at high velocities are investigated numerically. Current methodologies in the aircraft industry 
estimating water collection on aircraft surfaces are based on particle trajectory calculations 
and empirical extensions thereof in order to approximate the complex fluid-structure interactions. 
We perform direct numerical simulations (DNS) using the volume-of-fluid method in three dimensions, for 
a collection of drop sizes and impingement angles.
The high speed background air flow is coupled with the motion of the liquid in the framework of oblique stagnation-point flow. 
Qualitative and quantitative features are studied in both pre- and post-impact stages.
One-to-one comparisons are made with experimental data available from the investigations of Sor et al. (Journal of Aircraft 52 (6), pp. 1838-1846, 2015),
while the main body of results is created using parameters relevant to flight conditions with droplet sizes in the ranges 
from tens to several hundreds of microns, as presented by Papadakis et al. (AIAA Aerospace Sciences Meeting and Exhibit 0565, pp. 1-40, 2004). 
Drop deformation, collision, coalescence and microdrop ejection and dynamics, all typically neglected or empirically modelled, 
are accurately accounted for. In particular, we identify new morphological features in regimes below the splashing threshold in the modelled conditions. 
We then expand on the variation in the number and distribution of ejected microdrops as a function of the impacting drop size beyond this threshold. 
The presented drop impact model addresses key questions at a fundamental level, 
however the conclusions of the study extend towards the advancement of understanding of water 
dynamics on aircraft surfaces, which has important implications in terms of compliance to aircraft safety regulations.
The proposed methodology may also be utilised and extended in the context of related 
industrial applications involving high speed drop impact such as inkjet printing and combustion.

\end{abstract}

\begin{keyword}
drop impact, stagnation-point flow, spreading, splashing, volume-of-fluid, direct numerical simulations
\end{keyword}

\end{frontmatter}


\section{Introduction}
\label{sec:Intro}

Since the days of Worthington \cite{worthington1876forms}, the problem of droplet impact has offered the fluid dynamics 
research community exciting opportunities and challenges over the course of its history. 
For the first time in a systematic manner, in his book entitled \textit{A study of splashes} \cite{worthington1908}, 
Worthington makes use of early photographic technology (alongside careful 
sketchwork) to provide a comprehensive visual interpretation of splashing phenomena.
The framework has since captivated the interest of theoreticians and experimentalists alike, as it 
incorporates one of the most invitingly simple geometrical configurations,
while at the same time giving rise to diverse and rich phenomena of immense scope. 

A plethora of application areas benefit from understanding the outcomes of droplet impact events.
We emphasise in particular the role of droplet splashing (or absence thereof) in printing technologies (\cite{vanDam1,jung2012}),
combustion (\cite{Moreira1}), granular material interactions at all scales (\cite{thoroddsen2001,marston2012}), 
electronics (\cite{kim2007}) and spray-cooling in nuclear reactors (\cite{sawan1975}).
The design of superhydrophobic coatings in relation to droplet impact dynamics (\cite{Tsai2,Deng1}) 
is yet another prime example of the widespread applicability 
of this canonical problem.

Recent reviews provide an excellent insight into the state-of-the-art in the field within each decade (Rein \cite{rein1993} in the 1990's, 
Yarin \cite{Yarin1} in the 2000's and more recently Josserand \& Thoroddsen \cite{josserand2016drop}).
The area has witnessed a very strong surge in the past decade, fueled in part by the development of 
progressively more powerful imaging technologies, with both frame rates and resolutions capable of capturing 
details beyond the scope of previous equipment (see \cite{thoroddsen2008} as well). Furthermore, the 
improvement of numerical algorithms and usage of high performance computing has enabled computational studies 
that complement and inform both experimental and analytical work. We focus particularly on the volume-of-fluid 
package \textit{Gerris} \cite{popinet1, popinet2}, which is one of the most popular open-source tools due 
to its strengths in dealing with interfacial flows on a range of very different scales. Comparisons with experiments, 
as well as analytical work have been consistently robust, be it in cases of liquid-liquid impact \cite{Thoraval2,Agbaglah1} or 
impacts of liquid on solid surfaces \cite{Visser1,philippi2016drop,wildeman2016spreading}.

In the case of normal (perpendicular) impact at low-to-moderate velocities (and depending on specific fluid properties), 
an axisymmetric assumption can be used 
in analytical and computational investigations. The reduction in dimensionality 
is a significant advantage that has led to very efficient (axisymmetric) computations 
and good agreement with experiments. Visser et al. \cite{Visser1} for example, while 
innovating experimental technology enabling the time-resolved investigation of micron-sized drop impacts, have managed 
to conduct successful comparisons with direct numerical simulations at impact speeds of up to $50$ m/s, 
a regime which is commonplace in combustion, inkjet printing or aircraft-related applications.
In the respective scenario, the small drops spread onto the surface in what is known as pancaking motion, with 
the axisymmetric approximation remaining valid in the absence of splashing events.

In cases where spreading and later retraction rather than splashing occurs, the vast majority of efforts 
have been dedicated towards identifying quantities such as the maximal spreading radius 
(\cite{stow1981,clanet2004maximal,fedorchenko2005effect,Roisman2009,Schroll1} and most recently \cite{wildeman2016spreading}),
as well as the resulting minimal film thickness, retraction dynamics and the role of the internal boundary layer - see \cite{bartolo2005retraction} and 
in particular \cite{Eggers1} for a comprehensive investigation of the above.

At higher speeds however, there is still an ongoing debate as to how the splashing phenomena are first initiated, 
and the splashing threshold in particular. Up until the groundbreaking experimental investigation of Xu et al. \cite{Xu1}, 
there have been numerous attempts to characterise the transition from spreading to splashing dynamics in the classical impact problem in terms of drop-related 
parameters only (density, viscosity, impact velocity, surface tension). The Chicago group discovered, however, 
that decreasing the ambient air pressure may completely suppress splashing. As such, a host of additional modelling, experimental 
and numerical efforts have been initiated, with the work of Riboux et al. \cite{riboux2014experiments}
proposing a model deducing a threshold splashing velocity as a function of a generalised set of key parameters containing
the liquid density and viscosity, the drop radius, gas density and viscosity, the interfacial tension coefficient, 
as well as the nanometric mean free path of the gas molecules.
 
Once the drop splashes, there is very little attention dedicated to the ensuing dynamics, with the sizes and velocities 
of secondary drops being prohibitively small experimentally and computationally, although advances have taken place recently in terms of simplified models.
In particular, Riboux \& Gordillo \cite{riboux2015diameters} have proposed a one-dimensional approach to predicting sizes 
and velocities of ejected droplets for $\mathcal{O}(1)$ mm sized impacting drops and low speeds, finding reasonable agreement with experiments.

As underlined by Josserand \& Thoroddsen \cite{josserand2016drop}, there are several exciting challenges 
lying ahead, two of which are of great importance in the context of the present work. First of all, 
gaining an improved understanding of splashing, particularly in difficult high speed conditions of industrial relevance, is moving more and more within reach, 
and further detailed investigation using the available tools is needed. Secondly, oblique 
impacts are rarely analysed (exceptions being \cite{Mundo1,Sikalo1, bird2009inclined}) due to the additional flow complexity. 
Most often, qualitative rather than quantitative phenomena are explored in detail. The exceptions tend to focus on large 
scale effects at the level of the entire drop, as opposed to details at the level of the splashing itself and the interesting local structures arising.
Both of these themes lie at the heart of the present work, which focuses on the modelling 
and computation of oblique three-dimensional drop impact in aerodynamic conditions.

In aircraft-oriented research and design involving drop impact, the relevant scales are often dictated by the size of the parts 
that are most affected by phenomena such as water impingement, retention and finally icing and its prevention. The wings or nacelles 
are several metres long, while computing accurate air flows around them requires domains that span tens of metres in all dimensions. 
This becomes highly prohibitive in terms of accurate resolution of the intricate and sensitive physical effects 
pertaining to drop impact, which often happen at sub-micron scales in the order of tens to hundreds of microseconds. 
As such, particle-trajectory calculations of various degrees of complexity have thus far proven to be the only 
tractable solution in industrial setting.

There are several important limitations of current models, as pointed out by Gent et al. \cite{Gent1} in a relatively recent review:
\begin{itemize}
 \item droplets are assumed to be spherical and non-deformable as they approach the solid surface, hence topological transitions 
 such as the emergence of secondary drops either before or after impact are not considered;
 \item phenomena related to multiple drops such as collisions are completely ignored;
 \item aerodynamic drag, gravity and buoyancy are assumed to be the sole forces affecting the drop trajectories;
 \item whereas the local velocity of the air flow is embedded into the ordinary differential equations governing 
 the updates in drop trajectories, the liquid mass is assumed not to affect the surrounding air flow;
 \item once on the surface, empirical models translate the drop contribution towards liquid film formation 
 and its movement further downstream along the surface of interest.
\end{itemize}
Many of these assumptions become inaccurate in the context of the large supercooled droplets (larger than several 
tens of microns) found in the atmosphere. The difficulties outlined above have yet to be overcome, 
and most modelling is performed at a highly coarse-grained level 
(\cite{potapczuk1993ice,bragg1996aerodynamics,rutkowski2003numerical,wright2004,wright2005validation,wright2006,honsek2008,bilodeau2015}), with semi-empirical relations of varying complexity 
being proposed in order to match with the rich but ultimately limited experimental data available 
by NASA experiments conducted by Papadakis et al. \cite{papadakis2003,papadakisExp}.
The focus here is primarily on the final water retention values rather than the more fundamental problem 
of the detailed impact process, making it ideal from an engineering standpoint but offering limited 
insight into the underlying physics. In the past few years, the group at INTA/Madrid \cite{vargas2012mechanism,sor2015modeling} have looked 
in more detail into the deformation of large-scale drops prior to impact, with results 
that indicate regimes far more complex than captured by the typical assumptions mentioned above. 
Several studies focusing on recent numerical advances in the high speed regime ($>50$ m/s impact velocity) have emerged, 
particularly for impacts onto liquid, but also onto solid surfaces \cite{ming2014lattice, cheng2015numerical, guo2016investigation, Cherdantsev2017, Xie2017}.
These offer exciting opportunities to study short timescale phenomena beyond the reach of traditional particle methods, 
however up to this point there have been few attempts to integrate the drop impingement process 
into a framework that includes a more realistic model for the movement and effect of the air flow around the bodies of interest.

The present work bridges the relevant scales in the problem of drop impact onto aircraft surfaces 
and proposes a suitable model for the air flow around the solid bodies of interest in which we then 
accurately resolve the drop impingement process. While the drops are initialised as spherical 
sufficiently far away from the body, we characterise their deformation prior to impact and 
the spreading/splashing thereafter, depending on drop sizes and angles of impingement. 
We focus on the asymmetric features of the drop spreading when droplets are very small (less than a few tens of microns), 
phenomena which to our knowledge have yet to be observed. As the drop size increases, we quantify 
the sizes and positions of the secondary drops emerging as a result of the impingement 
and provide useful metrics for practitioners looking to improve water retention calculation methodology 
and a deeper understanding of the physics involved in the impact process under challenging conditions.
All flow parameters have been carefully chosen to match with previous experimental studies or known flight-specific values, 
while many of the quantified metrics are also compared to classical theoretical results where applicable.

The investigation is structured as follows. We introduce the proposed mathematical model in Section~\ref{sec:Model},
followed by a detailed description of the computational framework in Section~\ref{sec:Numerics}. 
We then analyse our findings in Section~\ref{sec:Results}, focusing 
on both pre-impact dynamics in subsection~\ref{subsec:preimpact} and post-impact dynamics in subsection~\ref{subsec:postimpact}.
These results are discussed and placed into context in Section~\ref{sec:Conclusions}, 
followed by concluding remarks.

\section{Mathematical Model}
\label{sec:Model}

In the present section we elaborate on how we adapt the classical problem of drop impact to the high speed flow conditions of interest 
around aircraft surfaces.
First we discuss some useful assumptions allowing us to reduce geometrical complexity in the problem in subsection~\ref{subsec:Scales},
after which we expand on the mathematical model itself, outlining the relevant equations, initial and boundary conditions.

\subsection{Scale Transition}
\label{subsec:Scales}

The full model geometry discussed in previous paragraphs (aircraft wings/fuselage components) is far too complicated - and specific - from many points of view. 
To begin with, our aim is to present a general methodology, applicable to a number of surfaces rather 
than a specific specialised geometry. Secondly, the multi-scale modelling of both the background air flow
around the large scale body and the splashing dynamics within the much smaller drop impact regions 
is beyond reach in terms of theoretical and current computational resources. 
We thus employ several simplifications 
to enable a closer inspection of a much more amenable problem, which still preserves the main physical 
characteristics we wish to address.

Based on the disparity between the two scales in the problem (the impacting 
droplet diameter and the solid body it impinges upon), we assume the curvature of the body 
to have negligible effects. 
To justify this approximation, the radius of curvature of the 
leading edge of a typical NACA airfoil or nacelle lipskin, the most sensitive regions to water retention and icing,
is estimated to be of $R_b = \mathcal{O}(10^{-1})$ m for standard commercial aircrafts.
For a reasonably large droplet 
of radius $R = 100\ \mu$m, we find $R/R_b \approx 10^{-3}$.
Thus, from the perspective of modelling the local droplet impact, the surface can be considered as approximately flat.
From a different viewpoint, we zoom in sufficiently close to the surface of the solid 
body, such that in the respective region the droplet diameter is the representative lengthscale 
and hence the details of the impact can be carefully studied.

\subsection{Governing Equations}
\label{subsec:Equations}

The framework of studying these fluids as incompressible in laminar flow conditions is a natural choice in the context of our problem, 
as the primary target flight regimes of take-off and landing are characterised by relatively low velocities compared to 
those reached at higher altitudes. Furthermore, most droplet impingement events are concentrated close to the leading 
edge of the geometries of interest, where the flow has yet to enter the transition from laminar to turbulent state. Even in such circumstances, 
a complex and likely empirical turbulence model would prevent the inspection of the detailed liquid dynamics, which is 
the main goal of the present investigation.

The model fluids are assumed to be incompressible, immiscible and viscous.
Subscript $1$ is used to refer to the fluid inside the drop (taken to be water),
whereas subscript $2$ decorates quantities in the surrounding (air) flow.
Let $\rho_{1,2}$ and $\mu_{1,2}$ denote the constant densities and dynamic viscosities of the two fluids in the system. 
The constant surface tension coefficient at the interface is given by $\sigma$.
Velocity vectors $\textbf{U}_{1,2} = (U_{1,2}, V_{1,2})$
and pressures $P_{1,2}$ are used in the formulation of the dimensional momentum and continuity equations
\begin{align}
 \rho_1 (\textbf{U}_{1t}+(\textbf{U}_1 \cdot \nabla)\textbf{U}_1) &= - \nabla P_1 + \mu_1 \Delta \textbf{U}_1, \label{eq:Droplets1} \\ 
 \rho_2 (\textbf{U}_{2t}+(\textbf{U}_2 \cdot \nabla)\textbf{U}_2) &= - \nabla P_2 + \mu_2 \Delta \textbf{U}_2, \label{eq:Droplets2} \\
  \nabla \cdot{{\textbf{U}}_{1,2}} &= 0. \label{eq:Droplets3} 
 \end{align}
 Gravitational forces are assumed to be negligible. 
 There are two lengthscales in the problem: the droplet diameter $D$, the natural choice for the reference lengthscale, and the size of the (finite) 
 computational domain $L$. We scale lengths by $D$, velocities by a reference background velocity $U_{\infty}$
 and pressures by $\rho_1 U_{\infty}^2$. The emerging non-dimensional parameters are
\begin{equation}
\textrm{Re}= \rho_1 U_{\infty} D/ \mu_1,\quad \textrm{We} = \rho_1 U_{\infty}^2 D/\sigma,\quad K = \textrm{We} \sqrt{\textrm{Re}} = \sqrt{\rho_1^3 D^3 U_{\infty}^5/ (\sigma^2 \mu_1)}.
\end{equation}
The Reynolds number $\textrm{Re}$ and Weber number $\textrm{We}$ appear directly from the non-dimensionalisation procedure, 
while the splashing parameter $K$ is introduced as an intrinsic element of a drop impact problem.
The expression, originally introduced by \cite{stow1981}, has been used to classify the possible outcomes of the impact. 
This parameter has been controversial in the literature and 
cannot independently account for the classification of the complicated impact process (see \cite{Xu1, mandre2012}), 
however it serves as an indicator of the force of the splashing and permits comparisons with 
previous investigations.

We also introduce density and viscosity ratios
\begin{equation}
r=\rho_1/\rho_2,\qquad m=\mu_1/\mu_2,
\end{equation}
and non-dimensionalise the governing equations~\eqref{eq:Droplets1}-\eqref{eq:Droplets3}, resulting in
\begin{align}
 \textbf{u}_{1t}+(\textbf{u}_1 \cdot \nabla)\textbf{u}_1 &= - \nabla p_1 + \textrm{Re}^{-1} \Delta \textbf{u}_1, \label{eq:Droplets4} \\ 
 \textbf{u}_{2t}+(\textbf{u}_2 \cdot \nabla)\textbf{u}_2 &= - r \nabla p_2 + rm^{-1}\textrm{Re}^{-1} \Delta \textbf{u}_2, \label{eq:Droplets5} \\
  \nabla \cdot{{\textbf{u}}_{1,2}} &= 0. \label{eq:dDroplets6} 
 \end{align}
 The non-dimensional timescale is $D/U_{\infty}$. 
 The typical fluid properties in the case of water and air at near freezing temperature (close to $0^{\circ}$ C) 
 are given as follows. Water has density $\rho_1 = 999.8\ \textrm{kg/m}^3$ and dynamic viscosity $\mu_1 = 1.16 \times 10^{-3} \ \textrm{kg/ms}$, while 
 the air density is $\rho_2 = 1.21\ \textrm{kg/m}^3$ and its dynamic viscosity $\mu_2 = 1.81 \times 10^{-5} \ \textrm{kg/ms}$.
 The constant surface tension coefficient is $\sigma = 7.2 \times 10^{-2}\ \textrm{N/m}$ and 
 a representative value for the velocity of the background flow is $U_{\infty} = 78.44 \ \textrm{m/s}$.
 This value has been selected to coincide with classical experimental investigations \cite{papadakisExp}, 
 as well as subsequent numerical investigations in the aerospace engineering community (e.g. \cite{bilodeau2015}).
 We underline the large density ($r=826.28$) and viscosity ($m=64.09$) ratios, which pose significant numerical challenges - these 
 are touched upon in Section~\ref{sec:Numerics}. Tables~\ref{tab:dropDeformationParameters} and~\ref{tab:dropImpactParameters}
 in the results discussion indicate the values of the key dimensionless groups in the problem and highlight 
 the violent high speed impact regime investigated here.

To define the interfacial conditions governing the motion of the drop, we 
assume a sharp interface $y=S(x,t)$; subsequently this is relaxed 
in the context of the volume-of-fluid methodology employed in the direct numerical simulations.
The prescribed interfacial conditions are, in order, the kinematic condition, the continuity of normal 
and tangential stresses, and continuity of velocity components:
\begin{align}
w_i &= S_t + u_i S_x + v_i S_y,\qquad i=1,2, \\
\left[\textbf{n}\cdot \boldsymbol{\mathcal{T}} \cdot \textbf{n}\right]^1_2 &= \textrm{We}^{-1} \kappa,\\
\left[ \textbf{t} \cdot \boldsymbol{\mathcal{T}} \cdot \textbf{n} \right ]^1_2 &= 0,\\
\left[\textbf{u}\right]^1_2 &= 0,
\end{align}
where $[(\cdot)]^1_2=(\cdot)_1-(\cdot)_2$ represents the jump across the interface, 
$\textbf{n}$, and $\textbf{t}$ are the unit normal and tangent to the interface, respectively,
and $\kappa$ is the interfacial curvature. The stress tensor $\boldsymbol{\mathcal{T}}$ is given by
\begin{equation}
 \mathcal{T}_{ij}=-p\delta_{ij} + \mu \left( \dfrac{\partial u_i}{\partial x_j} + \dfrac{\partial u_j}{\partial x_i} \right),
\end{equation}
where the appropriate subscript is used in different fluid regions. 
The initial and boundary conditions for the finite computational domain are described in the following subsection.

\subsection{Background Flow}
\label{subsec:Background}

One of the most important features of the model is the interaction 
between the liquid drop and the air around it. In typical experimental conditions, 
droplets are formed at the tip of an injection device and fall under gravity, with the height 
of the device being varied in order to adjust the terminal velocity and hence fix the relevant 
dimensionless parameters. In order to reach velocities beyond $\mathcal{O}(1)$ m/s it is necessary 
to have some form of ejection mechanism that ensures not only reproducibility of the shapes, but 
also a stability of the dynamics in early stages as the drop travels through the quiescent air flow 
and may become immediately sheared and violently deformed and broken up. As such, most of the 
investigations concerning velocities above $10$ m/s are restricted to very small drops (well below $100$ microns), 
such that surface tension is strong enough to preserve the approximately spherical shape of the drop.

In flight conditions, leading edge droplet impact can be locally embedded in a stagnation-point
flow which develops into boundary layers on either side of the geometry.
As such, most droplets encounter a developing boundary layer structure with a strong shear component. 
In an effort to reproduce the same type of air flow environment while preserving generality, 
we proposed an oblique-stagnation point flow model for the air flow, with the liquid drop 
being seeded sufficiently far away from the body on the dividing streamline of the flow. The reasons 
behind this choice are twofold:
\begin{enumerate}
 \item Far away from the surface the drop should retain its shape and setting a uniform velocity field 
 inside the drop with a zero (as in most desktop experiment setups) or purely horizontal (along the body) air motion would produce instantaneous 
 breakup of the drop. The choice for stagnation-point background flow and the initial position of the drop ensures that the air 
 flow undergoes only small changes until sufficiently close to the surface, which is when we expect the drop to start 
 deforming in real life conditions.
 \item The stagnation-point flow has the same characteristics in the near vicinity of the point of zero 
 velocities as on the aircraft surface, in that boundary layers are developing on either side of it and 
 growing as we move further downstream. As such the liquid drop is subjected to the shear flow naturally occurring 
 above the solid surface. This is best represented in the highly oblique impact cases,
 in which the air flow streamlines near the surface have strong deviations from their far-field orientation. Oncoming drops
 depart from their host streamlines close to the surface and their final impingement points 
 are well within the boundary layer growth region. The choice in initial positioning of the drop thus retains generality, while 
 ensuring suitable conditions for the early stages of the drop dynamics.
\end{enumerate}

There are however several points to be made prior to advancing to the mathematical description of the model. 
First of all, the dynamics of drops in uniform flow has been extensively investigated and the deformation 
characteristics for large enough drops are very rich (see \cite{Jalaal1} for a recent computational study). 
Therefore, even in the case of tailoring the initial position of the drop to a region of uniform air flow,
the drop is anticipated to suffer significant deformations as it moves towards the solid body. The size 
of the finite computational domain can then be used to alleviate (or enhance) this effect. Secondly, it should be 
noted that there still remains a fundamental difference to the practical scenario in which a solid body is moving 
through high liquid water content clouds (with stationary water drops of varying sizes) as opposed 
to drops impinging onto a static solid surface, as in the present case. Here we are enhancing the inertial 
contribution in the pre-impact drop dynamics and our choice in initial position of the drop does ultimately 
affect the liquid volume impinging onto the surface. Previous experimental results have however been used 
as guidance in order to best account for the complex flow dynamics, while retaining 
a suitable well-controlled flow environment.

With the above properties in mind, we underline that the background air flow poses its own non-trivial 
challenges. The history of the problem dates back to Hiemenz \cite{Hiemenz1}, who 
was the first to present a solution for the two-dimensional normal stagnation point flow. Howarth \cite{howarth1951cxliv} then 
extended the formulation to three dimensions.
The oblique case was first touched upon by Stuart \cite{Stuart1}, then later rediscovered independently by Tamada \cite{Tamada1} and Dorrepaal \cite{Dorrepaal1}.
There have been a number of corrections, extensions and generalisations on the main problem, such as \cite{wang2008similarity} and \cite{Tooke1}, including 
extension to two-fluid systems (air flow impinging onto liquid films above a solid surface), as studied by \cite{tilley1998oblique} and \cite{Blyth1}. 
As far as we know there is no general analytical solution to the three-dimensional stagnation-point flow problem at an arbitrary angle.
As such, we attempt to recreate this type of flow numerically using a combination of suitable boundary conditions that preserve 
its main characteristics in the interior of the domain.       

The three-dimensional computational box is selected to be of size $4L/D \times L/D \times 2L/D$ in $(x,y,z)-$directions, 
where $L/D$ is taken to be of size $20$, i.e. $20$ drop diameters. The flat solid surface is taken to be in the $(x,z)-$plane, 
with no-slip and impermeability conditions prescribed in this region, 
such that at $\textbf{u}_{1,2} = 0$ at $y=0$. 

In order to model the oncoming flow at an arbitrary angle of incidence 
$\theta_i$ 
we prescribe inflow conditions given by 
\begin{equation}
 u_2(x,L/D,t) = \cos(\theta_i), \qquad v_2(x,L/D,t) = -\sin(\theta_i) \textrm{ at } y= L/D.
\end{equation}

Laterally we impose typical free outflow conditions on all remaining four sides of the box.
The main reason for doing so relates to the movement of the secondary drops resulting as a consequence 
of the splash which cause perturbations in the flow field, making it difficult to fix velocities at the boundaries.
The initial conditions are set to 
\begin{equation}
 u_2(x,y,0) = \cos(\theta_i), \qquad v_2(x,y,0) = -\sin(\theta_i),
\end{equation}
prompting the need for the convergence of the background flow to a steady state prior 
to the inclusion of the liquid droplets into the computational domain. For all 
cases considered in the present investigation, an evolution of the flow spanning $100$ dimensionless time units
proved more than sufficient for this purpose, with a tolerance of $10^{-6}$ 
in the components of the velocity field selected to verify flow convergence to steady state.
We have confirmed this for all angles of incidence using a root mean square norm 
of the velocity field, presented in Fig.~\ref{fig:fig2}b. Time $t=0$ is taken to be 
the time at which the drop is seeded inside the domain and by this convention 
the direct numerical simulations begin at $t=-100$.

\begin{figure}[!hbtp]
\centering
\includegraphics[width=1.0\textwidth]{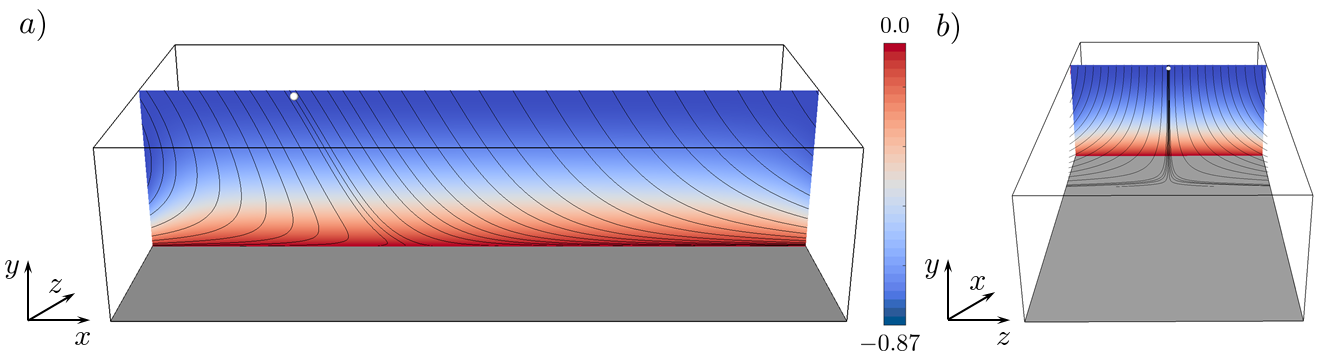}
\caption{Snapshots of the converged background velocity field obtained as a result of imposing uniform flow boundary conditions at an angle 
of $\theta_i = 60^{\circ}$ on the upper boundary, hitting a flat solid surface at the bottom, with outflow conditions on all lateral boundaries. 
The two cross-sections through a) the central $x-y$ plane and b) the $y-z$ plane illustrate the vertical velocity field (negative above, zero due to no-slip on the surface), 
as well as streamlines of the flow. The water drop, shown in white, is initialised on the dividing streamline near the upper border of the geometry.}
\label{fig:fig1}
\end{figure}
 
Focusing on the mid-$(x,y)$-plane (at $z=0$, see Fig.~\ref{fig:fig1}a), we find similar flow properties to the classical 
case of oblique stagnation point-flow in two dimensions.
Using the typical definition for the stream function $\psi(x,y)$, where the horizontal velocity component $u=\psi_y$
and the vertical velocity component $v=-\psi_x$, sufficiently far away from the wall the flow takes the form
\begin{equation}
 \psi(x,y) = k x y + \frac{1}{2} \zeta y^2.
\end{equation}
This is effectively a superposition of irrotational stagnation-point flow of strength $k$ and a uniform shear flow parallel to 
the solid surface (in the $x$-direction), where $k$ and $\zeta$ are scale constants (see \cite{Blyth1} for a recent exposition on this scenario). 
$\psi = 0$ denotes the dividing streamline 
onto which the liquid drop is initialised just below the upper boundary of the three-dimensional domain, with its center at $y = 19.25$ and $z=0$ 
and with $x$ varying as a function of the angle of incidence $\theta_i$ of the background flow.
Fig.~\ref{fig:fig1}a provides a visualisation of the converged flow field at the instance 
of the initialisation of the drop for the case when $\theta_i = \pi/3$.

We underline that, despite the flow being essentially two-dimensional in the upper part of the domain (below the inflow boundary), 
due to the presence of the solid surface and the lateral outflow condition, it develops its full three-dimensional structure 
close to the impingement region, with a single stagnation point being present in the flow irrespective of the impingement angle. 
This is best observed in Fig.~\ref{fig:fig2}a, but also in Fig.~\ref{fig:fig1}, where streamlines 
are drawn on top of velocity fields (illustrated in colour) plotted in different two-dimensional cross-sectional planes to indicate the deflection in the air flow.

%

\begin{figure}[!hbtp]
\centering
\includegraphics[width=1.0\textwidth]{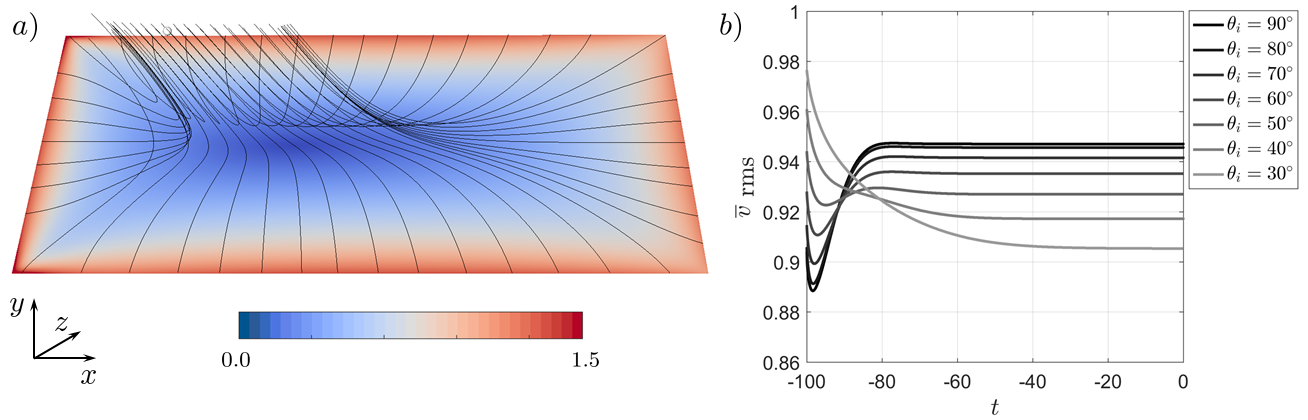}
\caption{a) Magnitude of velocity vector on an $x-z$ plane immediately above the surface, at $y=0.01$. A single stagnation point is visible in the center 
of the computational box, with streamlines aiding the visualisation of the flow as it increases in velocity towards the lateral boundaries. b) Root-mean-square 
norm of the velocity vector for different angles of incidence $30^{\circ} \le \theta_i \le 90^{\circ}$ of the background air flow.}
\label{fig:fig2}
\end{figure}

Once the background air flow has reached its steady state, the initially spherical liquid drop
is prescribed to enter the computational domain at a desired location $(x_i,y_i = 19.25, z_i=0)$.
The drop then inherits the local velocity field of the background air flow which is 
an approximately uniform flow directed towards the surface at an angle $\theta_i$, 
and is advected towards the solid surface.
The droplet shape is subject to physical deformations up to the time of its impact. 
Full hydrodynamic coupling determines its trajectory and shape, with no further assumptions being made beyond this point.

\section{Numerical Methodology}
\label{sec:Numerics}

The numerical computations in the present study have been carried out using the open-source package \textit{Gerris} \cite{popinet1,popinet2} 
(\href{http://gfs.sourceforge.net/}{http://gfs.sourceforge.net/}), which has been used extensively with great success by 
the multi-phase flow community over the last decade. The package is ideal for our purposes since it
accurately solves the incompressible Navier-Stokes equations (and a variety of additional multi-physics extensions) using 
the finite volume method and a volume-of-fluid approach to account for fluid-fluid interfaces. The schemes 
are second order accurate in both space and time, with strong adaptive mesh refinement capabilities ensuring 
the computational cost remains relatively low even in challenging multi-scale contexts such as those
in the present problem. 
In the following paragraph we elaborate on some of the specific 
measures used to ensure a good numerical performance, and also underline the
overall features of our extensive computational effort.

The large density ratio (recall that for water-air flows $r=826.281$) between the fluids may cause convergence issues 
for multi-grid Poisson solvers as the one used in 
\textit{Gerris} \cite{tryggvason1}, causing slow convergence or leading to a breakdown of the numerical solution altogether. 
A smoothing operator/filter has been proposed (\cite{popinet2,fuster2013}) in order to alleviate 
this. Spatial filtering consists of averaging over the corners 
of a computational cell (four in 2D and eight in 3D), which are in turn obtained 
by averaging the centered values of the corner neighbours. Applying the filter effectively 
smoothes the representation of the interface over a larger number of cells and can be applied 
any number of times, although previous investigations on drop impact argue 
that a single iteration of the filtering operator is sufficient \cite{thoravalThesis}.
As a result of this manipulation, the errors are maintained at a reasonable (and controllable) magnitude, while the convergence 
properties of the solver are much improved.

The qualities of the package, in particular in terms of adaptive mesh refinement (AMR), 
become evident in the study of the problem of drop impact at high velocities. The background 
air flow requires a strong level of refinement close to the surface of the solid body 
to account for the presence of the developing boundary layers around the stagnation point of the flow. 
At the same time, capturing the evolution of the fluid-fluid interface demands an appropriate 
resolution, enabling possible topological transitions. Splashing 
entails the creation and subsequent tracking of a large number of secondary droplets, which may 
or may not coalesce with other bodies of fluid. In addition, suitable choices 
for refinement with respect to sharp changes in vorticity were also implemented.
We also note the more stringent treatment required during touchdown, 
in which a reduced timestep and an extended local refinement region is necessary to avoid 
numerical artefacts.
The computational gain when comparing to the case of a uniform mesh is remarkable. 
The large domain would require $\mathcal{O}(10^{10-11})$ grid cells at the finest resolution,
however with the use of the adaptive mesh refinement this is decreased several 
orders of magnitude down to $\mathcal{O}(10^{6})$ degrees of freedom, which becomes significantly more tractable.
Many of the results presented would have reached considerable runtimes (as well as challenging 
memory and data storage requirements)
without the usage of adaptive mesh refinement, and possibly making many of the calculations presented here unrealisable.

We also employed the functionality to selectively eliminate droplets and bubbles whose dimensions are below 
a threshold number of grid cells (chosen to be $16$), thus fixing the minimum lengthscale that computations can account for.
Note that this is already well within the sub-micron scale. This feature becomes useful 
when secondary droplet break-off is violent and causes the fragmentation of the fluid 
into droplets of a very small size which suffer from geometrical reconstruction errors as a result of 
them spanning a small number of grid cells in each dimension. In practice the technique works by replacing 
the connected volumes (under 
a specified size) containing the drop fluid phase (water in our case) with the background fluid (air).
Furthermore, in our implementation complete droplet removal takes place if the droplets 
are found within one spatial unit of lateral boundaries in order to limit numerical artefacts 
when encountering the outflow region or sufficiently high above the surface of the solid 
body ($y>10.0$) to avoid high speed satellite droplets reaching the inflow boundary 
and causing numerical instabilities. In practice, the mentioned situation can be avoided 
by prescribing a larger computational domain that demands increased computational costs. 
The selective removal of droplets 
ensures that a geometry of manageable size can still be used reliably.
The flow in the vicinity of the impact region is unaffected by this treatment restricted to the near-boundary regions, 
hence no flow information is artificially lost. 

Many of the problems of interest require the treatment of a triple contact point 
between the solid surface and the liquid-gas interface. We note that the default mesh-dependent static contact 
angle model with a selected value of $90^{\circ}$ is used here. The limitations of this basic method, as well 
as proposed improvements have been recently discussed by
Afkhami et al. \cite{afkhami1}, who introduced a versatile dynamic contact angle model (implemented 
in an extension of \textit{Gerris}). 
In a related context, Pasandideh-Fard et al. \cite{Fard1} note that
the inertially dominated stages of the flow are unaffected 
by changes in the contact angle, which had been altered with the use of surfactants in their investigation.
In general, the suitability of the static contact angle model in the inertia-dominated 
spreading regime has been studied extensively (\cite{yokoi2009numerical,guo2016investigation}) and 
the present choice is not restrictive. We have experimented numerically 
with both grid sizes and different imposed static contact angle values in two dimensions, confirming 
that in the early stages of the impact we are in a regime which is insensitive to the choice of 
contact angle at the wall.

The runs in the present study have been performed at 
multiple resolution levels, varying from $2^{10}$ to $2^{12}$ grid cells per spatial 
dimension in each computational box. 
As the interfacial shape is set to be resolved at this level, this would translate to up to approximately 
$200$ cells per diameter for the initial spherical drop.
Before impingement we do not require such levels of refinement away from interfaces.
On the other hand, right before, during and after impingement, the entire liquid volume demands 
a strong refinement level. Despite these stringent requirements, with the chosen settings 
and drop sizes, each finest resolution cell spans from $0.097$ microns 
for the smallest impinging drops studied to $1.15$ microns in the case of the largest drops of initial diameter of just over $230$ microns. These levels 
have been selected to provide as much detail at the micron and sub-micron levels as possible. 
Many features, such as for example the minimal film thickness arising as a result of the spreading 
of a drop on the surface, have well-established sizes which are useful guidelines for
what scales need to be captured and are used for comparisons and validation.
From a more general standpoint, for the decision on a suitable level of refinement 
and mesh adaptivity setting we have relied on three main criteria: a) mass conservation; b) changes 
in defined metrics such as velocity and vorticity norms, secondary drop size distribution etc. and c) comparisons 
to available analytical predictions and experimental data in the literature. Once all three criteria have been met, 
the configuration in question was propagated towards full parameter studies over the variables of interest.
We emphasise that for the top two levels of refinement, volume conservation is accurate to within 
$1\%$ across the entire set of tests in the present work, with only the most challenging of test cases (the largest initial drop diameter prescribed) causing 
errors of the order of $3-4\%$ due to a combination of the difficult conditions (high Re/We) and the selective drop removal mechanism introduced above,
with smaller scale features being more frequent in this scenario.
A typical computation under these conditions requires in excess of $2\times 10^{3}$ CPU hours for the lowest resolution
tested and approximately $10^4$ CPU hours for the more refined cases. 

In the next section we describe, in turn, our computational results for pre-impact deformation and post-impact dynamics, 
concentrating on both fundamental phenomena and aspects related to the larger scale system itself.

\section{Results}
\label{sec:Results}

Once the drop is initialised in the steady background flow, it travels towards the surface 
guided by an initially uniform (but $\theta_i$-dependent) velocity field, with streamlines deforming as the solid surface is approached. 
Analysing the deviation from the initially spherical shape as a function of time is 
one of the primary goals of the present work, since, as noted earlier, this effect 
is often overlooked in standard water retention calculation models. 

In order to provide a suitable 
validation framework for the present results, we have tailored the parameters to coincide 
with a subset of the data of the only experimental investigation of the pre-impact deformation and break-up phenomena we are aware of - see
Vargas et al. \cite{vargas2012mechanism} and Sor et al. \cite{sor2015modeling}. 
Therein, an experimental setup consisting of a monosize droplet dispenser, 
a rotating arm with a model wing fixed at its end, 
as well as associated motor and camera equipment are used to capture the drop dynamics as 
the solid body approaches the liquid droplets at velocities of up to $100$ m/s. As a result of the very violent 
regime, the size variation for the drops is restricted to $D=300\ \mu$m and above. Very few pixels 
per drop diameter are visible below this threshold and the resulting images can no longer 
be comprehensively analysed. As a consequence, 
in the results that follow we have selected three values within the respective range, as well as one smaller drop, 
typical of the sizes found in the high liquid water content regions aircraft travel through.
The drop sizes, as well as all other associated dimensionless parameters are 
summarised in Table~\ref{tab:dropDeformationParameters}, where we underline 
that we have used a reference velocity of $U_\infty = 90$ m/s (the same as in the main series of experiments \cite{sor2015modeling}) 
and the physical properties of water and air at relevant near freezing temperatures.

\begin{center}
\captionof{table}{\label{tab:dropDeformationParameters} Relevant dimensionless parameters in the case of pre-impact deformation studies 
in high speed conditions, matching in drop diameter to a subset of the studies performed by Sor et al.\cite{sor2015modeling}.}
\begin{tabular}{|c|c|c|c|c|c|c|}
\hline
 $D$ [m]& $\textrm{Re}= \rho_l U_{\infty} D/ \mu_l$ &  $\textrm{We} = \rho_l U_{\infty}^2 D/\sigma$ & $\textrm{Oh} = \sqrt{\textrm{We}}/\textrm{Re}$& $\textrm{Ca} = \mu_l U_\infty/\sigma$ & $\textrm{St} = \mu_g/(\rho_l D U_\infty)$  \\  \hline \hline  
 $128 \times 10^{-6}$ & $8653.717$ & $10936.183$ & $0.012$ & $1.263$ & $1.803 \times 10^{-6}$ \\
 $362 \times 10^{-6}$ & $24473.794$ & $30928.893$ & $0.007$ & $1.263$ & $6.376 \times 10^{-7}$ \\
 $634 \times 10^{-6}$ & $42862.943$ & $54168.282$ & $0.005$ & $1.263$ & $3.640 \times 10^{-7}$ \\
 $1048 \times 10^{-6}$ & $70852.309$ & $89539.999$ & $0.004$ & $1.263$ & $2.202 \times 10^{-7}$ \\
\hline
\end{tabular}
\end{center}

Following impact itself, depending on the relevant parameters, the drop is anticipated to either spread due to its momentum 
and subsequently recede under surface tension effects, or, in the cases of the larger drops, to splash and 
break up into secondary droplets which move away from the surface but may later re-impinge. 
Droplets found in the atmosphere typically lie within the interval of $20-250\ \mu$m in diameter 
and as a consequence water catch studies reported in the literature \cite{papadakis2003,papadakisExp,wright2005validation,wright2006,honsek2008,bilodeau2015}
are found in this regime. 
We consider four test cases ($D=20, 52, 111, 236\ \mu$m) for complete analysis of pre- and post-impact dynamics,
in order to facilitate comparisons with results in the field and provide further insight under 
flight conditions of practical interest. The complete list of parameters is provided in 
Table~\ref{tab:dropImpactParameters}, where the same water-air configuration is used, 
however this time with reference velocity $U_{\infty} = 78.44$ m/s, in agreement with 
datasets discussed in previously mentioned studies.

\begin{center}
\captionof{table}{\label{tab:dropImpactParameters}  Relevant dimensionless parameters in the case of long-time drop impact direct numerical simulations
in high speed conditions, matching in median volumetric diameter to a subset of the studies performed by 
Papadakis et al. \cite{papadakis2003}. 
The splashing parameter $K = \textrm{We} \sqrt{\textrm{Re}}$ varies between $ 6.283 \times 10^4$ and $2.547 \times 10^7$.
}  
\begin{tabular}{|c|c|c|c|c|c|c|}
\hline
 $D$ [m]& $\textrm{Re}= \rho_l U_{\infty} D/ \mu_l$ &  $\textrm{We} = \rho_l U_{\infty}^2 D/\sigma$ & $\textrm{Oh} = \sqrt{\textrm{We}}/\textrm{Re}$& $\textrm{Ca} = \mu_l U_\infty/\sigma$ & $\textrm{St} = \mu_g/(\rho_l D U_\infty)$  \\  \hline \hline  
 $20 \times 10^{-6}$ & $1352.143$ & $1708.779$ & $0.031$ & $1.263$ & $1.154 \times 10^{-5}$ \\ 
 $52 \times 10^{-6}$ & $3515.573$ & $4442.824$ & $0.019$ & $1.263$ & $4.438 \times 10^{-6}$ \\
 $128 \times 10^{-6}$ & $8653.717$ & $10936.183$ & $0.012$ & $1.263$ & $2.079 \times 10^{-6}$ \\
 $236 \times 10^{-6}$ & $15955.291$ & $20163.588$ & $0.009$ & $1.263$ & $9.779 \times 10^{-7}$ \\
\hline
\end{tabular}
\end{center}

For the smallest $20\ \mu$m drops we consider an extensive parameter study in terms of impingement angles $10^{\circ} \le \theta_i \le 90^{\circ}$ 
in increments of $10^{\circ}$. This enables a detailed analysis of the effects related to the competition between inertial and capillary regimes, 
while noting the influence of the background flow on the drop dynamics. For the more challenging larger droplets we focus on two 
specific cases, namely $\theta_i = 60^{\circ}$ and $\theta_i = 90^{\circ}$, guiding us towards results in both 
symmetric and asymmetric impact, described in full detail in subsection~\ref{subsec:postimpact}.

\subsection{Pre-impact dynamics}
\label{subsec:preimpact}

In the present subsection we describe qualitative and quantitative features related 
to the motion of droplets prior to them impacting the solid surface. 
Intuitively we expect the most deformation 
and possible break-up to happen close to the solid surface as the air flow slows down and the droplet 
encounters developing boundary layers. We note however that, particularly for large drops, 
a rich dynamics characterised by so-called bag break-up and rupture can be observed 
even in the case of simple uniform flow and in the absence of any streamline deflection \cite{Jalaal1}.
These strongly time-dependent morphological changes underline the importance of one of 
the parameters in the presented model, namely the initial position of the drop relative to the solid surface. 
If prescribed too far away from the surface, the initial spherical drop may become completely fragmented 
by the time it reaches the surface, while seeding it too close to the surface may not allow sufficient 
time for its natural dynamics to occur before impingement. As such, the comparison to the experimental results 
from INTA/NASA \cite{vargas2012mechanism, sor2015modeling} serves as an important validation step.
The authors focused on describing 
and modelling the change in shape, as well as the consequences thereof in terms of predicting 
the drag coefficient of the evolving shape. They found that for moderate-sized droplets (with diameters in the hundreds 
of microns) the approximation of the shape as an oblate spheroid proves to be reasonably accurate, 
quantifying this deformation as $a(t)/R$, where $a(t)$ denotes the evolving 
major semi-axis of the spheroid, normalised by the initial drop radius. This value was reported 
to increase smoothly from $1.0$ as the drop is sufficiently far away from the surface to 
values in the range of $1.3$ for $D=362\ \mu$m, to $2.0$ for $D \approx 1$ mm, increasing monotonically as a function 
of the size of the initial drop. As they approach the surface,
the larger drops suffer considerable deformations in which the symmetric framework postulated before 
is no longer applicable. Finally, when close to within $10$ mm of the solid surface, the drops violently 
break up into a cloud of secondary droplets which can only be described qualitatively in the experiments.

\begin{figure}[!ht]
\centering
\includegraphics[width=1.0\textwidth]{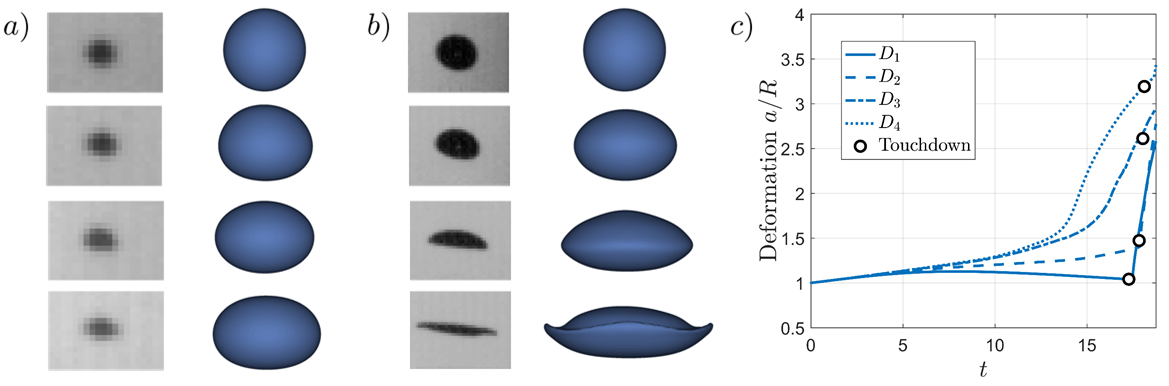}
\caption{Pre-impact drop deformation visualisation for spherical drops of diameter a) $D=362\ \mu$m and b) $D=1048\ \mu$m. 
Inside each panel the left images are experimental results by Sor et al \cite{sor2015modeling}, while the right images are the
corresponding DNS results. The images are reproduced with permission by Instituto Nacional de T\'{e}cnica Aeroespacial.
c) Quantification of the drop deformation in terms of the drop semiaxis $a$ normalized by the initial radius $R=D/2$, with the corresponding parameters 
described in Table~\ref{tab:dropDeformationParameters}. The timestep 
at which the drop first touches the solid surface is also highlighted with an open circle.}
\label{fig:fig3}
\end{figure}

Example evolutions of the droplet shapes are shown in panels a) and b) of Fig.~\ref{fig:fig3}, 
in which we analyse the deformation of a relatively small drop ($D=362\ \mu$m), 
as well as a large drop ($D=1048\ \mu$m) alongside their experimental counterparts.
In the former case, we find that the proposed mild deformation into an oblate spheroidal 
shape is recovered and good qualitative agreement with the experiments is found. The same 
applies for the latter larger drop case, in which the flattening of the shape is much more pronounced 
and asymmetric features arise in the latter stages. Note how the center of gravity 
of the shape shifts towards 
the lower part of the drop in the third subimage, only to develop secondary structures around the 
edges which ultimately rupture from the main shape and break off into smaller droplets prior to impact.
It should be noted that there is a difference in timescales when comparing the experimental 
and computational results; in the experimental data the deformation takes place 
over a distance of several hundred drop diameters, whereas in all computational results this evolution 
takes place within the prescribed distance of roughly $20$ initial drop diameters. The flow field 
and its extensional nature is effectively scaled down to the size of the computational box.
  
From a quantitative perspective, for comparison purposes we use the same semiaxis deformation metric $a(t)/R$ 
in Fig.~\ref{fig:fig3}c to uncover an excellent agreement with the experimental data. We mark the time of impact 
with an open circle and note that the obtained values are within $10\%$ of their experimental counterparts, 
while the evolution of this measurement in time also shows the same features. Notably, for the larger drop 
we plot the full extent of the liquid volume (accounting for the shedding of secondary droplets). 
If these are to be excluded, at a distance of half a diameter above the solid surface, the deformation 
is found to be $1.36$, $1.72$ and $1.85$ for the $362\ \mu$m, $634\ \mu$m and $1048\ \mu$m drops, respectively,
with approximately $1.3$, $1.7$ and $1.94$ being the equivalent values in the experiment. The inclusion 
of secondary drops becomes visible around $t \approx 15$ in both cases 
and causes an increase in this metric to just below $2.0$ and $3.0$ for the two largest droplets, 
indicating the complexity of the flow in the respective regimes as the drops approach the surface.

The computational framework developed here can be used to access information on the flow field 
and drop shape at distances very close to the surface that are beyond the frame-restricted capabilities
of current powerful video technology. 
Consequently, we consider the case 
of a smaller drop of initial diameter $D=128\ \mu$m and find very small deviations 
from the imposed shape during its entire evolution. A small initial flattening of the shape into an oblate spheroid 
suffers corrections prior to impact and ultimately impinges almost undeformed. 

\begin{figure}[!h]
\centering
\includegraphics[width=1.0\textwidth]{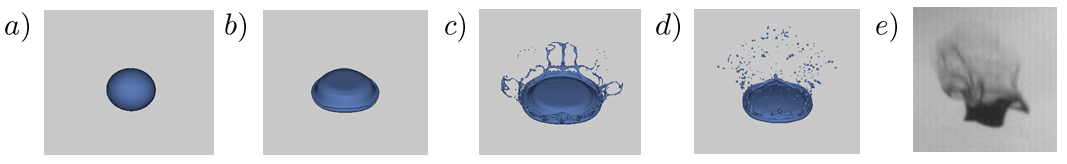}
\caption{Initially spherical droplets of diameter a) $D=128\ \mu$m, b) $D=362\ \mu$m, c) $D=634\ \mu$m, and d) $D=1048\ \mu$m, at the moment of impact 
onto a flat solid surface, having been deformed by the background stagnation point flow. The smallest drop retains its shape, 
while the edges of the largest drop break up into a large number of secondary droplets even before impact. This compares favourably 
to e) previous experimental investigations of drop deformation prior to impacting a moving solid body ($D_{\textrm{exp}}=1048\ \mu$m as well) by Vargas et al \cite{vargas2012mechanism}. 
The last image is reproduced with permission by Instituto Nacional de T\'{e}cnica Aeroespacial.}
\label{fig:fig4}
\end{figure}

For completeness, all four cases are illustrated in Fig.~\ref{fig:fig4} at the last computed timestep before touchdown, with the last 
$D=1048\ \mu$m case being placed side-by-side with its experimental counterpart. 
For the smallest drop, deformation is hardly visible 
(as confirmed by Fig.~\ref{fig:fig3}a), with an approximately spherical liquid volume impinging onto the surface. A strong 
flattening of this shape with the beginning of breakup features becoming visible around the edges, takes place for slightly larger drops 
and this ultimately leads to progressively smaller liquid fragments/drops being shed from the sides. In the largest drop volume case, 
the cloud of droplets behind the main liquid volume becomes visible and resembles the experimental result.
  
In what follows we focus on the impingement process itself and in particular on the spreading or splashing 
characteristics of the flow, as well as the associated secondary drop formation and dynamics.

\subsection{Post-impact dynamics}
\label{subsec:postimpact}  

\begin{figure}[!ht]
\centering
\includegraphics[width=0.9\textwidth]{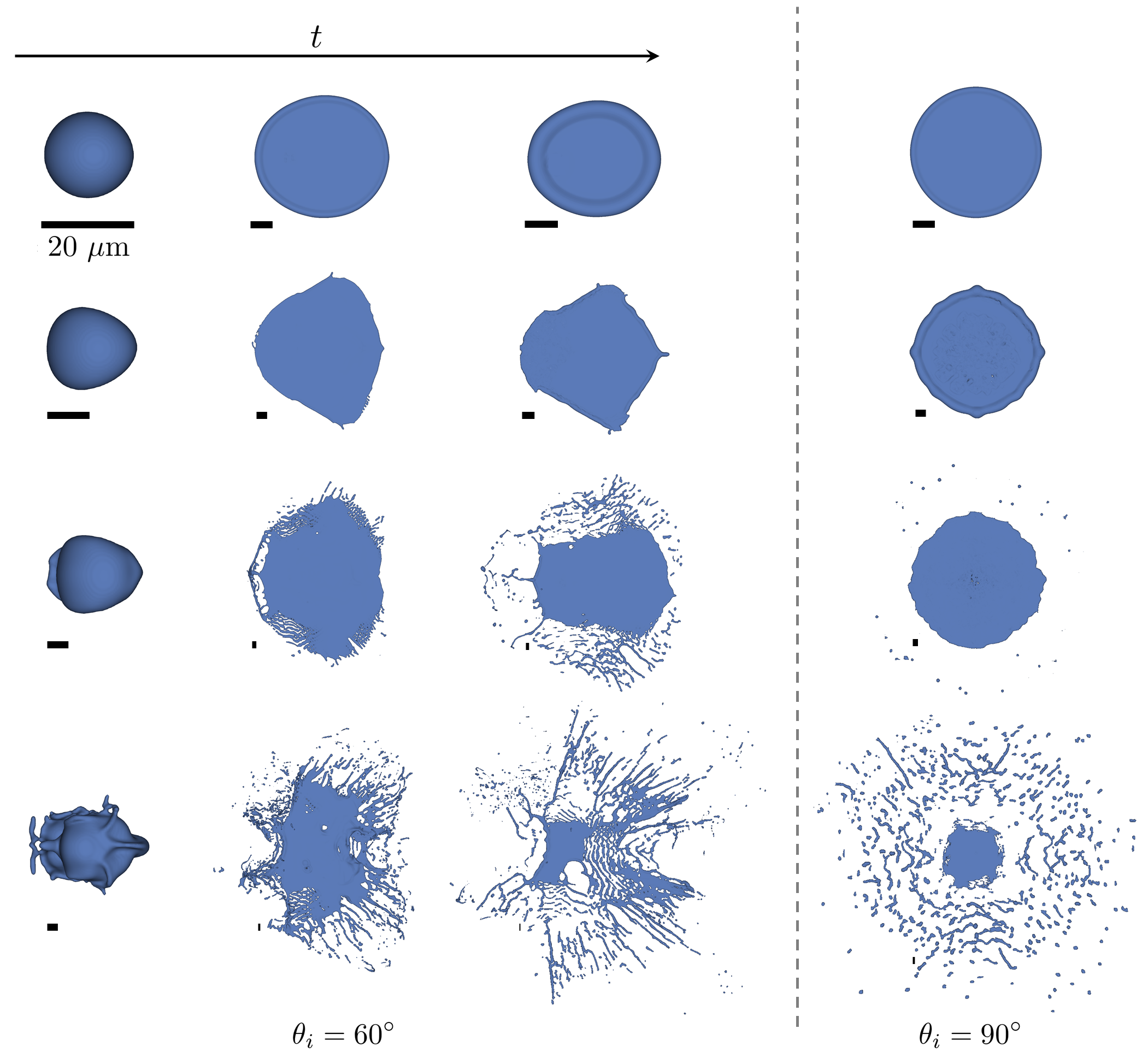}
\caption{Splashing dynamics for drops of sizes $D = 20,\ 52,\ 111$ and $236\ \mu$m (each row represents a different drop 
size, with the complete list of parameters defined in Table~\ref{tab:dropImpactParameters}) at an angle 
of incidence of $\theta_i = 60^{\circ}$. The left column illustrates the drop shapes as their center of mass is at $y=D$ above the surface, 
the images in the second column are plotted at the dimensionless timestep at which the 
drop illustrated on the top row (smallest drop, with initial diameter $D=20\ \mu$m, and relatively regular spreading behaviour)
reaches its maximum spread $t_s^{\textrm{max}}$, while the third column shows 
the drop shapes ten time units later, once either retraction or more pronounced splashing has occurred. The rightmost column is used 
to visualise the splashing for the $\theta_i = 90^{\circ}$ impact case at $t_s^{\textrm{max}}$.}
\label{fig:fig5}
\end{figure}

Once the drop approaches the region very close to the wall, the gradually thinner 
air film below is forced to move away laterally and the pressure underneath the droplet continues 
to grow until impact takes place. We note the presence of either a single or multiple 
air bubbles entrained under the surface. In the classical context with a quiescent air flow 
and small to moderate impact velocities, the size and evolution of the air bubble is well studied, 
and its effect on the splashing process itself has been shown to be negligible \cite{riboux2014experiments}.
In the present case however, there are two fundamental differences from the traditional impact problem 
due to the very high impact velocity, as well as the strong pre-impact drop deformation, 
particularly in the oblique impingement cases. In the first instance and on the basis of Fig.~\ref{fig:fig5}, 
we will provide a qualitative assessment of the results. The studied parameter space consists of 
the two cases outlined in Table~\ref{tab:dropImpactParameters} 
of normal impact and oblique impact at $60^{\circ}$, and four different drop diameters, producing eight 
studies in total. The time sequence of top views of the liquid drop on the left hand side of Fig.~\ref{fig:fig5}, illustrates the drop 
shape at 
three key times 
in its evolution, namely:
\begin{enumerate}[i.]
 \item when the center of mass of the drop lies one initial diameter above the surface (left column);
 \item when the drop reaches its maximum spread on the surface and before retraction under capillary forces takes place (middle column);
 \item ten time units later, which serves as an indication of how the longer timescale of the impact develops 
 into either retraction for the smaller drops or violent rupture and splashing for the larger drops (right column).
\end{enumerate}
For the $90^{\circ}$ impact case we concentrate on the second of the above time instances, namely when the smallest 
drop reaches its maximum diameter - results are shown in the right hand side column of Fig.~\ref{fig:fig5}. 
In each image a reference 
lengthscale of $20\ \mu$m is added as a visual aid to the extent of the drop atomisation (or lack thereof).

The smallest drop size (initial diameter $D=20\ \mu$m) impingement is characterised by  
inconsequential pre-impact deformation with the approximately spherical shape retained 
up to very near the time of impact, followed by a strong spreading motion in a highly inertial regime, 
finally followed by retraction due to surface tension. Intriguing corner-type features emerge particularly 
for the oblique impact cases due to the directionality of the impact, which will be discussed in detail 
in subsequent paragraphs.
Referring to the oblique impact scenario, the asymmetry becomes more visible for medium sized drops (at the order of $100\ \mu$m in initial diameter) 
prior to impact and particularly after impact as fluid volumes have sufficient momentum to 
overcome surface tension and push outside the typical nearly circular contour, instead spreading laterally outward 
towards the front of the drop. The dynamics is however still dominated by one large fluid volume 
from which small secondary drops are ejected as the drop increases in size. We underline that the imposed angle of incidence 
has a clear influence on the angle and extent of lateral spread of the liquid mass.
The largest drops ($D=236\ \mu$m) experience violent splashing, with visible liquid threads forming in the forward and laterally outward directions 
as the fluid mass disintegrates into hundreds of droplets. 
Similar features are observed in the normal impact case in terms of fragmentation, with 
traditional spreading motion transitioning to azimuthal instabilities, followed 
by a rupture of the liquid rim into small drops, but with a main fluid mass still intact near the impact site.
Ultimately a violent fragmentation breaks up the liquid volume into thin filaments near the surface, 
and numerous secondary drops are advected away from the impact region under the influence 
of the background flow.

Conducting a systematic analysis of the drop's morphology during the early and intermediate 
stages of the impact is most accessible for the smallest drops (below several tens of microns in initial diameter, 
top row of Fig.~\ref{fig:fig5}, when no splashing occurs),
where early and very recent analytical results are available for comparison when
$\theta_i = 90^{\circ}$. Following this baseline, the generalisation to the predominantly three-dimensional 
effects of the asymmetric impact are best constructed. Even in the normal impact case however, the presence of 
the non-quiescent air flow at high speeds is anticipated to produce some modifications in the standard metrics 
surrounding the characterisation of the impingement process, which will be emphasised in the following paragraphs.

\begin{figure}[!hbtp]
\centering
\includegraphics[width=1.0\textwidth]{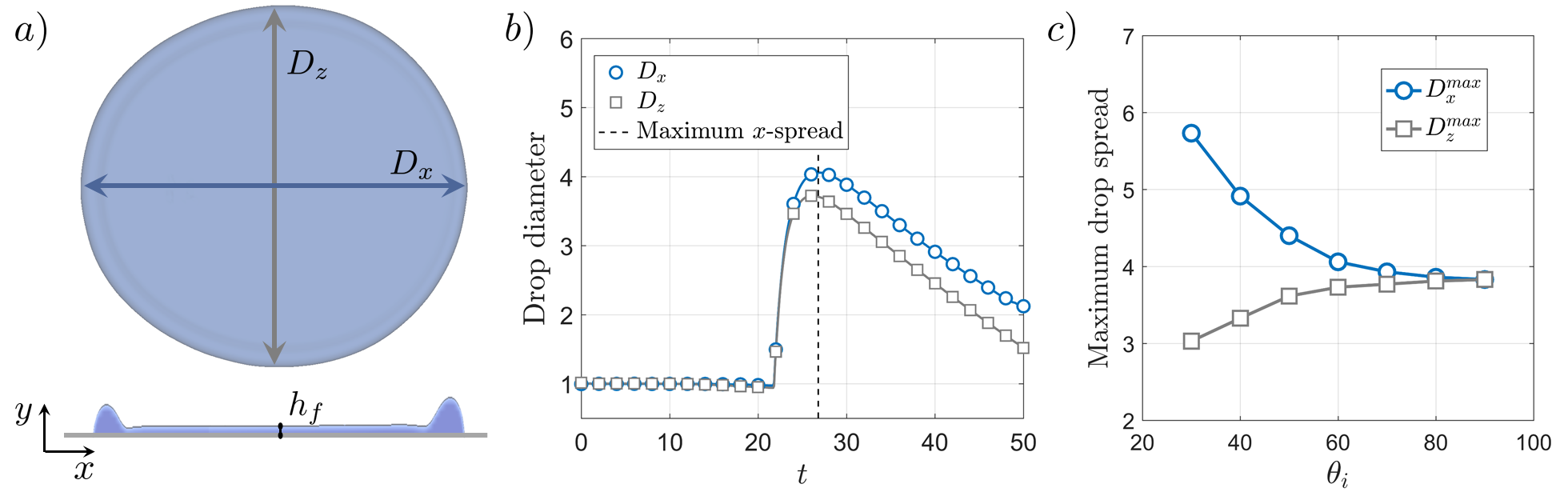}
\caption{Spreading dynamics of microdroplets with an initial diameter of $D = 20\ \mu$m at angles of incidence ranging from $30^{\circ}$ to $90^{\circ}$.
a) Top view schematic of the spreading diameter in the $x-$direction (the impingement direction) and the $z-$direction, as well 
as the minimal film thickness $h_f$ (in side view below), for an angle of incidence of $\theta_i = 60^{\circ}$. 
b) Evolution in time of the spreading and later retracting liquid drop for the $60^{\circ}$ impingement 
angle case. c) Summary of the maximum spread in both $x$ and $z$ for a collection of angles of incidence, indicating the transition from 
symmetric spreading to a strongly asymmetric final shape in the direction of impact.}
\label{fig:fig6}
\end{figure}

In order to aid future comparisons, in Fig.~\ref{fig:fig6} we define several quantities of interest, namely 
the time-dependent drop diameter 
in the $x-$direction $D_x(t)$ (the direction of impact for the non-normal incidence cases), 
the drop diameter in the transverse $z-$direction $D_z(t)$, as well as the height of the drop near 
its center of mass $h_f$. The first two metrics are best observed from the top view ($x-z$ plane) presented in the 
top part of panel a), while a cut through the $x-y$ plane provides information on the minimum thickness of the film.
The entrapment of a small air bubble due to impact cushioning, results in a small variation 
in the drop's curvature just above this feature, which is why for the relevant local minimum 
we select a point where this local adjustment is negligible. 
Several notable studies (see Introduction) have addressed the topic of the maximum spread $D_m$
of the drop in normal impact conditions, with the recent investigation of Wildeman et al. \cite{wildeman2016spreading} 
chosen as reference here. Plugging our parameters into their main result, we find $D_m \approx 4.112$, 
which compares very well with the computed value for $\theta_i = 90^{\circ}$ in Fig.~\ref{fig:fig6}c.
Symmetry in this case is preserved and we find $D_x^{\textrm{max}}  = D_z^{\textrm{max}} \approx 3.9$, which alongside the good agreement 
also indicates that the surrounding flow has a limited influence on the maximum spread. 

As the angle of incidence is decreased down to $\theta_i = 30^{\circ}$, the flow speed in conjunction with 
the increasingly pronounced directionality of the impact 
enables the liquid mass to advance towards the front side (in the $x-$direction) of the impacting drop, 
pushing more strongly towards the front edge and increasingly distorting it in this direction. Fig.~\ref{fig:fig6}c 
indicates this monotonic increase in $D_x^{\textrm{max}}$ and decrease in $D_z^{\textrm{max}}$ as 
$\theta_i$ is reduced, with the final aspect ratio being measured at almost a factor of two. 
We point out that in this regime the drop is also subjected to a stronger air flow as it lies 
further away from the dividing streamline and the background flow velocity has an increased magnitude. 
For illustrative purposes, in Fig.~\ref{fig:fig6}b we expand on how the maximum diameter values 
are obtained in the asymmetric cases, with the two diameters $D_x(t)$ and $D_z(t)$ being shown throughout 
their evolution for an angle of incidence $\theta_i = 60^{\circ}$. The dynamics in the $x-$direction 
is chosen as reference, as this is the dominant motion due to our choice in impact directionality. 
The value of $D_z^{\textrm{max}}$ is then defined as the value of $D_z(t_x^{\textrm{max}})$, 
where $t_x^{\textrm{max}}$ it the timestep at which $D_x$ reaches its maximum, despite 
it not necessarily being the highest absolute value in the $z-$direction. The figure 
shows negligible deformation up to the time of impact $t \approx 20.0$, followed by a sharp increase 
in diameter in both directions but more strongly in $x$, with the rim finally retracting under 
the effect of surface tension from all directions. The reference values (see vertical dashed line) 
derived from similar studies of each incidence angle are then used to construct Fig.~\ref{fig:fig6}c.

Another key morphological metric we consider is the minimum film height, as extracted near 
the drop center, sufficiently far away from the entrapped bubble. 
In the normal impact case and in the strong inertial regime described here, 
Eggers et al.\cite{Eggers1} estimate this thin film height to reach a minimum $h_f/R \approx \textrm{Re}^{-2/5}$, 
which would give $h_f \approx 0.028$ in our case. This is the height at which the thinning film reaches the liquid boundary 
layers within the drop itself and ceases its decrease. The result obtained in our 
investigation is $h_f \approx 0.033$ and we found no evidence of significant 
variation as a result of modifying the angle of incidence. The very slight 
overestimation is perhaps counterintuitive given that the fast air flow pushing from above 
would be expected to enhance the thinning effect. We note that even at these small 
lengthscales the mesh is sufficiently fine with several gridpoints spanning 
the thin film region; changes in the resolution did not result in meaningful 
changes of this value.

\begin{figure}[!h]
\centering
\includegraphics[width=1.0\textwidth]{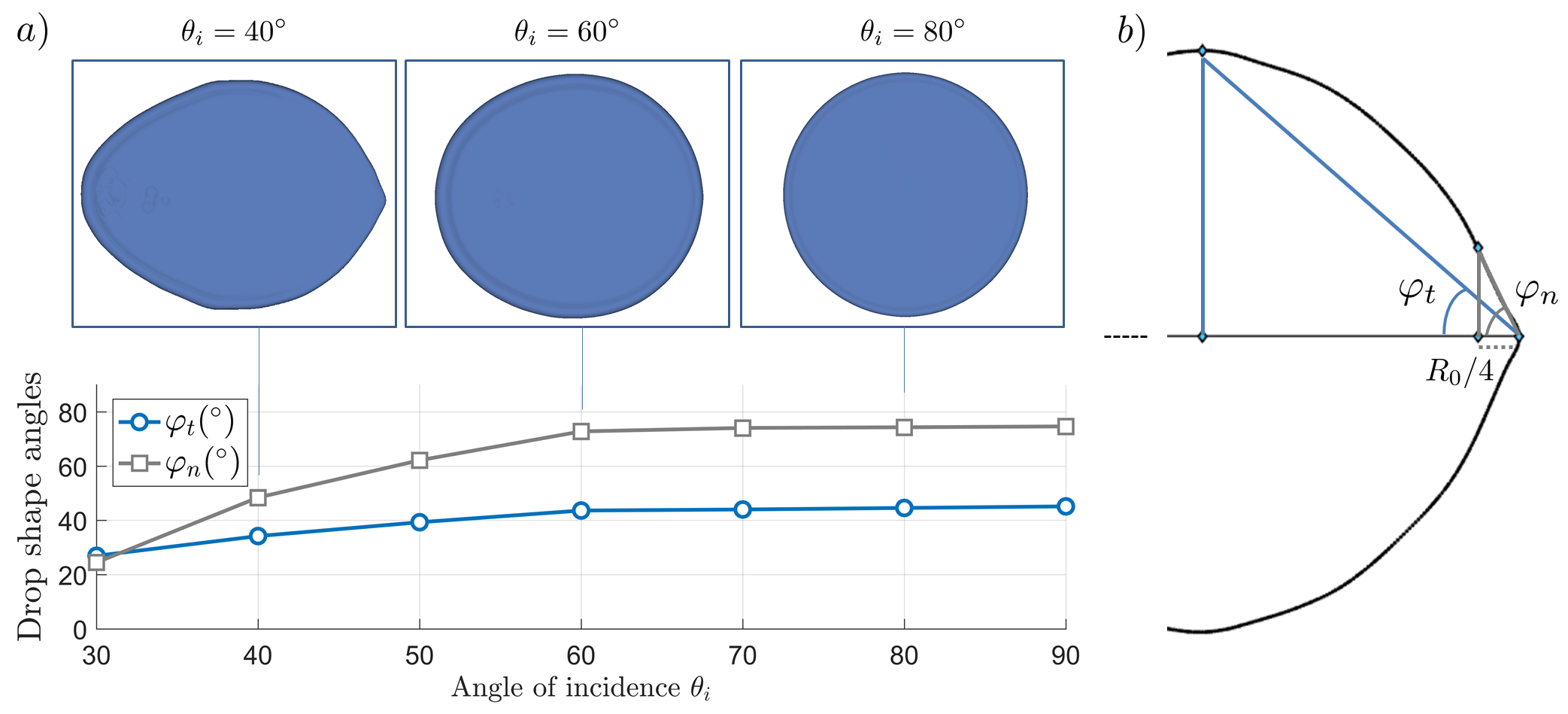}
\caption{a) Characterisation of the geometric feature arising at the leading (front) side of the drop due to the oblique impact. 
Angle $\varphi_t$ is measured from the most advanced point of the drop in the direction of impact to the maximum in the spread 
in the perpendicular direction of the same plane, while $\varphi_n$ represents the more local feature arising at $0.25R_0$ behind the 
front, where $R_0$ represents the initial drop radius. Both angles are defined in panel b), while the three insets present top views 
of the drop shape at the moment of maximum spread in the $x-$direction, the timestep at which all the angles in the figure are calculated.}
\label{fig:fig7}
\end{figure}

One of the most salient features of the drop impact in the modelled high speed regime is the emergence 
of a corner-type feature near the advancing front of the spreading liquid mass; this feature 
becomes highly prominent, particularly as the 
angle of incidence $\theta_i$ is $60^{\circ}$ or lower. 
Above the respective angle, normal impact is characterised by approximately axisymmetric behaviour,
while in slightly oblique impacts ($\theta_i \approx 70^{\circ}-80^{\circ}$) the footprint can 
 be described as elliptical, although a slight symmetry-breaking tilt to the front becomes observable 
 on the lower side of this range.
A comprehensive analysis 
of the corner-type property has been performed for angles varying in the range $10^{\circ} \le \theta_i \le 90^{\circ}$
and small drop size (initial diameter $D=20\ \mu$m), and the results are given in Fig.~\ref{fig:fig7}. 
Herein we track the evolving drop shape from above, and concentrate on the moment where its spreading 
diameter reaches its maximum value for each of the particular cases. 
Two angular metrics are then defined as illustrated on the right hand side of Fig.~\ref{fig:fig7}:
$\varphi_t$ (angle from the tip of the advancing front in the impingement $x-$direction to the top part of the drop, the maximum in the $z-$direction) 
is a more global measure of the deformation, whereas $\varphi_n$ is a local measure 
of the corner angle near the tip of the advancing front, defined by a triangle whose base is fixed to be a quarter 
of the initial radius $R/4$, as shown in the figure.
We emphasise that while the discussed feature is called a corner (or of corner-type) throughout this subsection, 
the shape would be more accurately described as an apparent corner, since locally near the tip of the advancing 
front surface tension always induces a smoothing of the shape.

The progressively more stretched shape of the drop, as well as the evolution near the tip of the advancing front capturing 
the corner-type feature itself are both embedded in the above quantities, which are presented at 
the bottom of Fig.~\ref{fig:fig7}, with examples of the underlying drop shapes depicted in the row above for $\theta_i = 40^{\circ},\ 60^{\circ}$
and $80^{\circ}$. 
In the intermediate case small distortions of the liquid rim are already visible, while 
at $40^{\circ}$ a pronounced outgrowth near the advancing front
selected by the direction of impact is clearly identifiable.
Due to the preserved axisymmetry, at $\theta_i = 90^{\circ}$ we compare the numerical 
results with simple predictions. 
We naturally expect $\varphi_t \approx 45^{\circ}$ and 
based on the maximal spreading radius described in Fig.~\ref{fig:fig6}c, the definition of the angle $\varphi_n$, as well as using basic trigonometry,
we estimate $\varphi_n \approx 75^{\circ}$. We recover $\varphi_t = 44.93^{\circ}$ and $\varphi_t = 74.76^{\circ}$ by analysing the data, 
which is well aligned with the anticipated axisymmetric evolution.
Both angle measurements are expected to decrease in oblique impact scenarios, with 
the elongation of the liquid shape gradually reducing their values as $\theta_i$ decreases. 
This is indeed 
the case, with a smooth monotonic variation in $\varphi_t$ finalising at approximately $22^{\circ}$ for the $\theta_i = 30^{\circ}$ impingement case.
The local angle $\varphi_n$ naturally begins at a much higher value, but again, as the impingement angle $\theta_i$ decreases, 
the deformation of the spreading drop is more is enhanced and this results in a steady drop from $\varphi_n \approx 75^{\circ}$ 
for the normal impact case down to $\varphi_n \approx 25^{\circ}$ at $\theta_i = 30^{\circ}$. The slope characterising 
this decrease becomes markedly larger in absolute value below $\theta_i = 60^{\circ}$. 
Note however that due to the shape irregularity this definition becomes less practical for angles below $\theta_i = 50^{\circ}$,
as high variation is induced depending on the choice of distance defining $\varphi_n$.
In particular, the top left hand side inset in Fig.~\ref{fig:fig7} reveals the formation of a small 
finger-like extension that becomes smoothed out under surface tension. The strongly varying curvature 
of this shape near the tip makes it difficult to design a universally useful local metric to describe the corner,
which is why the choice of a distance of $R_0/4$ behind the advancing tip should in some sense be interpreted with 
an embedded variation, as would any other choice.

Finally, a physical interpretation of the corner formation process is proposed as 
a combination of two different mechanisms, one related to the background flow, the other 
to the liquid movement itself. In the latter case, during the spreading motion 
liquid pushes into the front of the rim in a preferential direction given by the progressively 
more oblique impact as defined by $\theta_i$. There is sufficient inertia to drive more liquid mass 
towards the advancing front, however this is insufficient to overcome surface tension in the lateral direction 
due to the relatively small size of the drops. As such, the liquid that accumulates in the rim in the lateral regions is also steered towards 
the front of the drop, where at the meeting point the conditions for a localisation of the interfacial 
shape into a corner are met before surface tension relaxes this feature. At the same time, 
high speed air is pushing from above in the same direction as the primary impacting motion, further guiding liquid into this 
front region. This becomes far more evident for the lower impingement angles in which the drop spreads 
on the surface in a region several diameters away from the flow stagnation point and as such 
the local shear forces become gradually stronger, supporting the complete manifestation 
of the observed corner at the advancing front of the spreading drop.

\begin{figure}[!ht]
\centering
\includegraphics[width=1.0\textwidth]{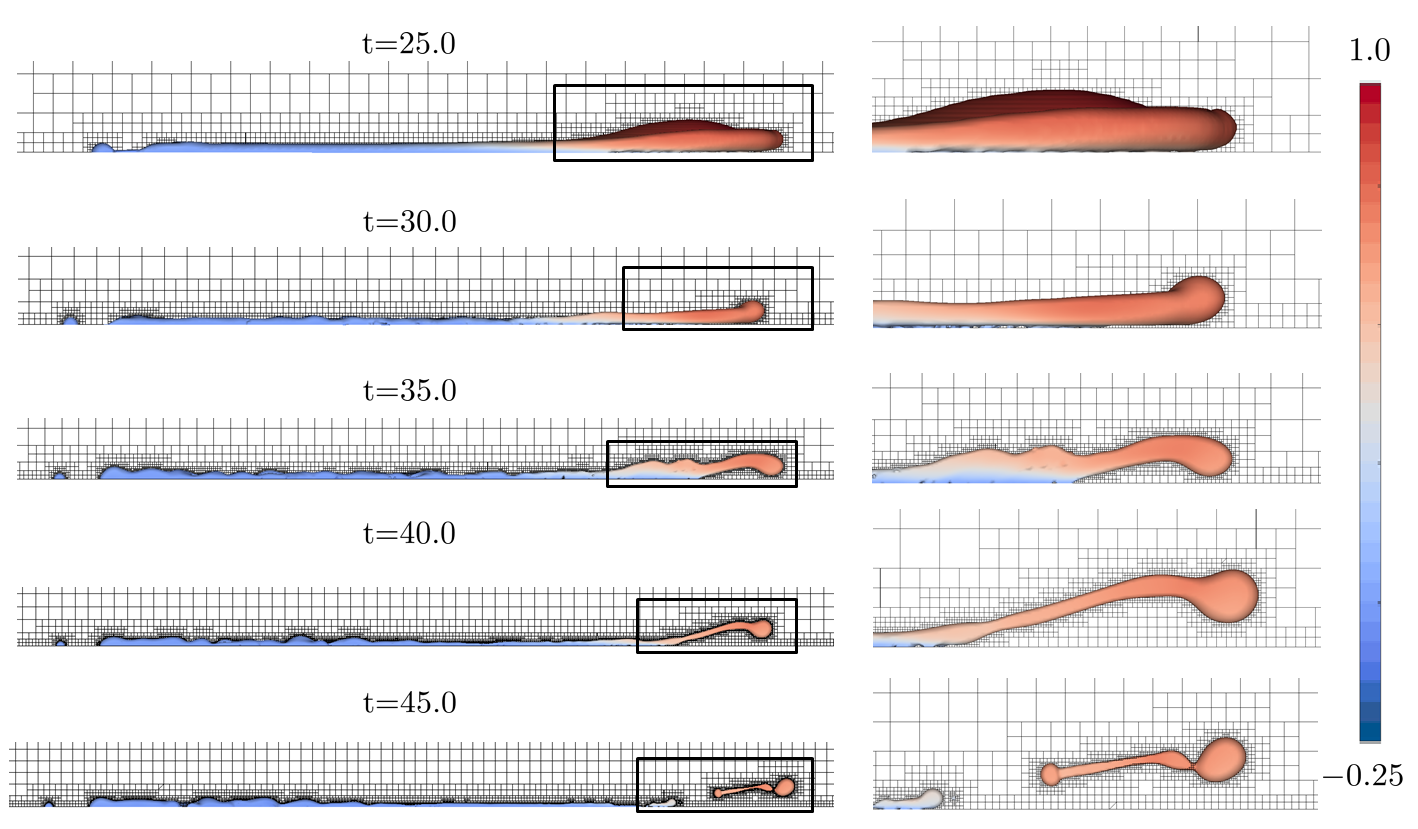}
\caption{Side view ($x-y$ plane) of the late time impingement dynamics of a $D = 20\ \mu$m drop at an angle of $\theta_i = 10^{\circ}$, 
resulting in the break-up of a liquid volume 
in the leading region. The panels on the right 
illustrate a magnification of this area.
The interface is coloured in the magnitude of the horizontal velocity at the respective points, while the adaptive grid underlying 
each timestep is also shown. For reference, the smallest grid cell measures approximately $0.39\ \mu$m in dimensional terms.}
\label{fig:fig8}
\end{figure}

For impingement angles lower than $\theta_i \approx 30^{\circ}$, we find entirely different phenomena 
captured in Fig.~\ref{fig:fig8}. The near-glancing 
incidence of the drop, coupled with a more uni-directional background air flow,
both contribute to a pinch-off near the advancing front of the liquid volume, as opposed 
to the creation and subsequent relaxation of a corner-type feature. For most of 
its development, the region near the front of the drop advances on top of a very thin liquid film, 
with the bulk of the liquid mass eventually catching up with significant horizontal velocity (used 
to colour the liquid interface in each subfigure in Fig.~\ref{fig:fig8}). The advancing front of the strongly 
elongated shape retains sufficient inertia to eventually detach from the surface and subsequently 
break off into several small liquid drops that progress at very high speed towards the edge of 
the finite computational domain, with the majority of the fluid quantity retained on the solid surface.
One of the additional causes underlying the observed fillamentation
is that the drop lands further away from the global stagnation point in the background flow, implying that locally
the flow is predominantly moving in the direction of the spreading in the front, promoting the lateral movement as opposed to 
pushing down onto the liquid.

\begin{figure}[!h]
\centering
\includegraphics[width=1.0\textwidth]{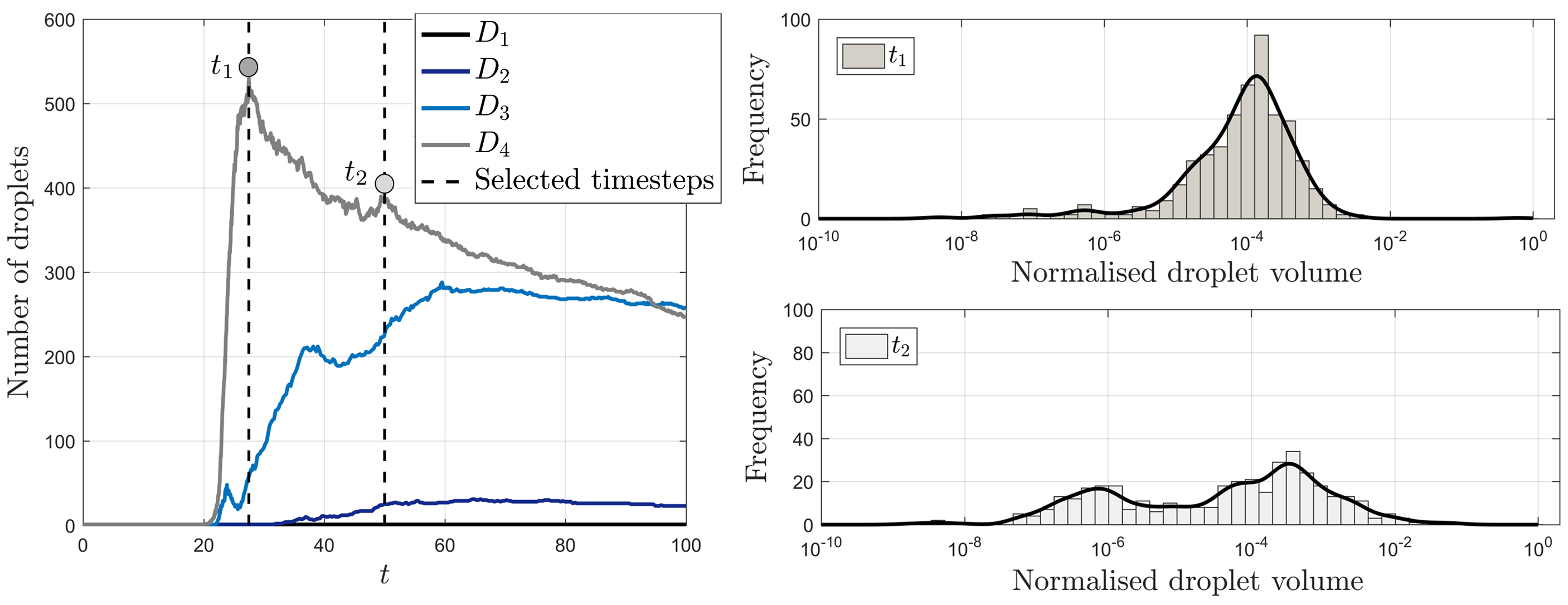}
\caption{Secondary droplet ejection characteristics as a result of spherical drops of initial diameter $D = 20,\ 52,\ 111$ and $236\ \mu$m impinging
onto a solid surface at an angle of incidence $\theta_i = 60^{\circ}$. Evolution of the number of droplets in time for each case (left), with the two panels 
on the right indicating the secondary drop size distribution (normalised by the initial droplet size) for the $236\ \mu$m drop at the two different times, $t_1$ and $t_2$, highlighted in 
the left panel.}
\label{fig:fig9}
\end{figure}

Having discussed at length the rich features already appearing in the impact of the smallest drops,
we proceed to a quantitative study of the splashing dynamics throughout 
the entire duration of the direct numerical simulations for the full range of drop sizes considered; these results are 
summarised in Fig.~\ref{fig:fig9}. Recall that the drops are initialised at $t=0$, 
impacting the surface at $t \approx 20.0$ and engaging in either spreading motion or generation of secondary droplets 
being tracked over $80$ subsequent dimensionless time units. Detailed results are presented for
the asymmetric case with an angle of incidence $\theta_i = 60^{\circ}$ and four different 
droplet sizes ranging from $20\ \mu$m to $236\ \mu$m.

Focusing on the left panel of Fig.~\ref{fig:fig9}, we notice the effect of initial drop size 
on the formation and break-up of secondary drops. The smallest 
drop follows the described spreading and retraction motion detailed in subsection~\ref{subsec:postimpact}
and no secondary drops are ejected from the surface. Instead, a steady spherical cap 
solution is observed during the final stages. At intermediate sizes, but still below 
an estimated $100\ \mu$m threshold, small drops detach from the edges of the rim 
in all directions however with a stronger preference towards the direction of impact, 
with approximately $20-30$ secondary droplets being swept away by the air flow and 
advected towards the lateral outflow boundaries; a few re-impinge 
onto the surface and remain at such respective locations.
For the largest droplets of initial diameter $D=111\ \mu$m and $D=236\ \mu$m,
a very violent splashing motion ensues (also visible in Fig.~\ref{fig:fig5}), 
with several hundreds very small drops moving away from the impingement region. 
As expected, due to the larger drop size the effect of surface tension is weaker and 
break-up into progressively smaller liquid fragments is more pronounced. 
In such cases, even the secondary droplets are subject to subsequent break-ups, with 
a minimum size being again restricted by surface tension.
We have made extensive verifications of the selected grid refinement and minimum cell size 
in order to restrict numerical artefacts at this level. In fact, the final resolution 
for these studies was partly selected in light of a convergence to a minimal secondary
drop size captured by our simulations. We note, however, that the selective droplet removal 
procedure performed primarily to avoid instabilities around the boundaries, does affect 
the secondary droplet count, with some of the smallest droplets ejected as part of the prompt splash away 
from the surface at very high speed being removed from the finite computational domain 
within several time units after the impact.

The size and position of each individual drop is tracked after impact and hence 
statistical information on the secondary drops is compiled and studied dynamically. 
Two particular points in time have been selected for visualisation purposes on the right hand side
of Fig.~\ref{fig:fig9}, which represent the early post-impact stage when the maximum 
number of secondary drops is found in the domain ($t_1$), followed by the point in time halfway through 
the evolution at $t=50.0$ when the main body of fluid no longer ejects secondary drops in the impact region and 
the secondary drops are airborne ($t_2$). Both 
of these apply to the case of the largest $D=236\ \mu$m initial diameter drop, 
selected due to the impact and splash producing the richest secondary drop dataset.

\begin{figure}[!h]
\centering
\includegraphics[width=1.0\textwidth]{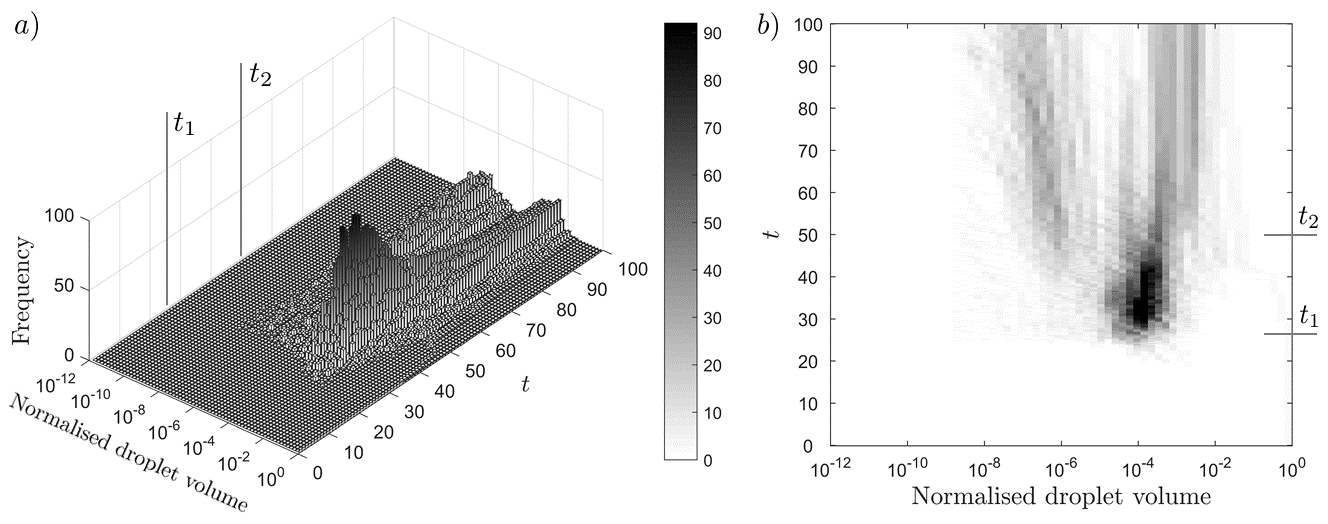}
\caption{Secondary droplet size distribution evolution characteristics for the largest drop 
in the test batch summarised in Table~\ref{tab:dropImpactParameters} and indicated as $D_4$ in Fig.~\ref{fig:fig9}, impinging
onto a solid surface at an angle of incidence $\theta_i = 60^{\circ}$. 
a) Left panel: three-dimensional view, with timesteps $t_1$ and $t_2$ detailed in Fig.~\ref{fig:fig9} marked with vertical bars.
b) Right panel: top view of the same dataset in the form of a contour plot.}
\label{fig:figR1DistributionEvolution}
\end{figure}

Once the impact has taken place and sufficiently many secondary drops have formed, 
the volumes of these liquid fragments follows an approximately log-normal distribution (see detail at $t_1$),
centered around a mean of $10^{-4}$ relative to the volume of the initial drop. If assumed to be spherical 
(which is seldom the case) this translates into droplets with a radius of $1/20$ relative to the radius of the initial drop.
As further fragmentation takes place due to the interaction between the fast movement of the drop 
and the surrounding air boundary layer flow, a second local maximum becomes prominent, 
with tens of drops with volumes of the order of $10^{-6}$ relative to the initial volume being present 
far away from the impact area. Under the action of surface tension, these drops are often 
more regular (spherical) in shape if still airborne, with some of them re-impinging 
far away from the impact region and becoming spherical caps as in the case of the previously 
studied $D=20\ \mu$m initial diameter drops. In fact, the distribution highlighted 
at $t_2$ in Fig.~\ref{fig:fig9} also includes the minimum droplet volume captured within 
this computation, which is found to be of just less than $10^{-7}$ or of a radius of $1/200$ relative 
to the initial drop - just above $1\ \mu$m in dimensional terms (recall the initial drop diameter is $D=236\ \mu$m in this 
numerical experiment).

The full evolution of the secondary drop size distribution in this case 
is illustrated with two different visualisation techniques in Fig.~\ref{fig:figR1DistributionEvolution}.
We are concentrating on the case of angle of incidence $\theta_i = 60^{\circ}$, however 
we have found the qualitative behaviour described below to be consistent with variation in $\theta_i$.
The observed log-normal distribution evolves (with the number of drops increasing, but the structure 
being retained) over roughly $15$ time units. At this stage ($t \approx 50$), most drops in the system follow one of two 
general tendencies:
\begin{enumerate}
 \item the larger drops in the distribution are the ones which are detached from the main liquid mass 
 but lie on the solid surface (after early dynamics or later re-impingement) as approximately spherical caps. 
 After they reach this configuration they will only increase in size as a result 
 of coalescence with neighbouring spherical caps or incoming smaller secondary drops that re-impinge onto the surface.
 This region, the right hand side local maximum centered around $10^{-3}$ in normalised volume) remains relatively steady in both number of drops and extent of variation.
 \item the fragmentation process around the left hand side second local maximum in the distribution (with mean of approximately $10^{-7} - 10^{-6}$) 
 is rich and spans roughly 
 two orders of magnitude in normalised volume. These are primarily airborne drops that 
 continue to travel, break up or coalesce as a result of the interaction with the background air flow.
\end{enumerate}
Rather remarkably, we find that the separation between the two types of drops (the two local maxima in the distribution)
 is relatively well preserved when changing both initial drop size and angle of impingement. 

 The same type of data analysis as presented above may be useful in furthering our understanding 
 of the dynamics very close to the splashing threshold, with possible comparisons 
 with the work of Riboux \& Gordillo \cite{riboux2015diameters} becoming an interesting 
 future line of investigation. In its current form, the high speed regime and presence 
 of the background flow in the model make such direct comparisons difficult, however 
 early calculations indicate that the first drops ejected as part of the splashing 
 process lie encouragingly close to the previously mentioned predictions.

\begin{figure}[!h]
\centering
\includegraphics[width=1.0\textwidth]{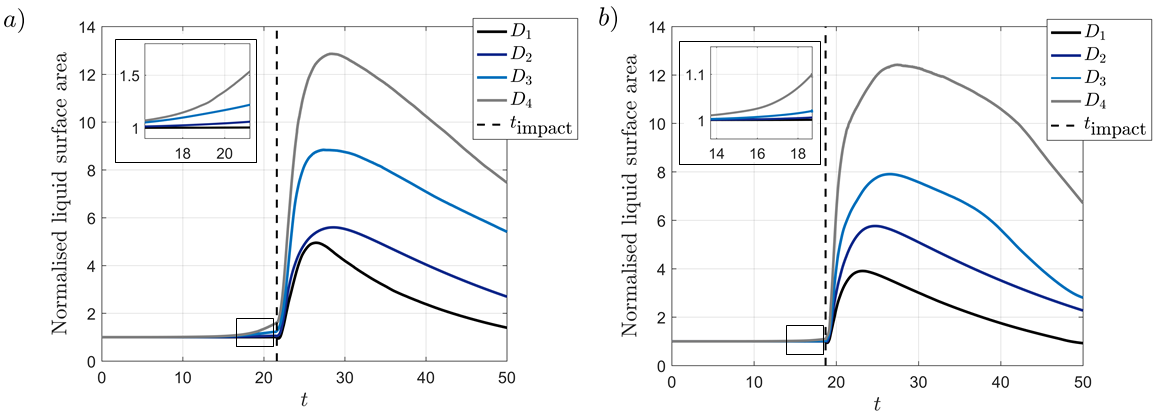}
\caption{Normalised liquid-gas surface area as a function of time 
for four different drop sizes (summarised in Table~\ref{tab:dropImpactParameters}) and
two impingement angles: a) $\theta_i = 60^{\circ}$ and b) $\theta_i = 90^{\circ}$.
The insets in each subplot concentrate on the evolution of the surface area during the five 
time units just before the moment of touchdown, marked by a vertical black dashed line in the full scale images.}
\label{fig:figR1SurfaceArea}
\end{figure}

The present work was in part motivated by the question of how extending our understanding of liquid droplets impinging onto 
solid surfaces in high speed conditions acts as one of the early building blocks in the broader context of icing 
prevention and aircraft safety and design.
As a result of the deformation and splashing dynamics forming the subject of the present work,
a useful metric to discuss is the evolution of the liquid-gas surface area,
which may in the future be considered in view of coupling to thermodynamic effects.
We illustrate our findings for the four cases introduced in Table~\ref{tab:dropImpactParameters}
and two angles of incidence, $\theta_i = 60^{\circ}$ and $\theta_i = 90^{\circ}$ in Fig.~\ref{fig:figR1SurfaceArea}.
To aid the discussion around the quantitative information, we introduce two simple approximations 
of what could be anticipated in light of previously elucidated dynamics and focus on the normal impact scenario for clarity.

For the smallest drops studied ($D_1$ here), we find that spreading and retracting behaviour is still characteristic 
and as such we can use a straightforward analogy in terms of a flat cylinder of appropriate radius and height to estimate 
the corresponding surface area. In particular, considering the time of maximum spread $t_s^{\textrm{max}}$, we find 
a value of the resulting $D_x^\textrm{max} \approx D_z^{\textrm{max}} \approx 3.9$ (see Fig.~\ref{fig:fig6}c).
Assuming (and having verified that) volume conservation holds, we find an approximate height for the cylinder, 
which in this case is roughly $1/24$. As we do not account for the bottom of the cylinder (the side adhering 
to the solid surface), the adjusted formula for surface area (accounting for the top surface and the side) 
gives a normalised result $S_1^{\textrm{max}} \approx 3.83$, in 
excellent agreement with the corresponding maximum in Fig.~\ref{fig:figR1SurfaceArea}b.
Now looking to the other end of the spectrum i.e. the largest drops considered in which case strong splashing is observed, the drop size distributions
in Fig.~\ref{fig:fig9} indicate that most of the drops have volumes with means of 
either $10^{-3}$ or $10^{-6}$ relative to the initial drop volume. Accounting again for conservation of volume, 
we would anticipate the normalised surface area to vary between $10$ and $100$ should we naively assume all drops to be airborne
and perfectly spherical. Instead we of course have a combination of these drops, many of which are also
on the surface (with the area in contact with the surface not counted), effects which would 
all bring our estimate closer to the former value of $S_4^{\textrm{max}}  \approx 10$. For both impingement angles 
we notice that this estimate does not deviate significantly from the obtained value.

Comparing the surface area evolution for both scenarios, we naturally find an increase 
with drop diameter, however this increase appears to be much more ordered in the normal impact case in 
terms of both distance between maxima and profile shapes. It should also be noted that, as the insets present,
the pre-impact deformation and hence surface area generated is more pronounced for the case with oblique incidence.
Once touchdown occurs, the initial strong increase in surface area is dominated by the spreading mass of liquid 
in both cases, with fragmenting secondary drops only adding negligible features to the evolution for large initial diameter cases.
Once a maximum is reached, capillary forces cause smooth retraction for the smaller drops. This 
is in contrast to the larger drop cases, in which, despite the apparent smoothness of the curves, a closer inspection reveals prominent bumps indicating 
individual fragmentation or coalesence events for small drops. Ultimately capillary retraction as well as 
drops exiting the finite computational domain through lateral boundaries lead to a decrease in overall liquid-gas surface area.

While much of our attention is dedicated to modelling and quantifying 
aspects related to single drop impingement dynamics at high speeds, we emphasise that the produced datasets 
encoding sizes and locations in space and time of all secondary drops becomes 
central in determining further re-impingement events and elucidating retention properties in general.
From a broader perspective, throughout the present section we have described not only fundamental flow features, 
but also intricate dynamics and a level of physical detail which is invariably omitted once highly simplified coarse-graining 
procedures are considered instead.
In a practical context, this information may provide an accurate physical foundation for engineering models on larger scales
and is the subject of ongoing work in our group.


\section{Conclusions}
\label{sec:Conclusions}

In the present work, three-dimensional drop impact at high velocities has been investigated numerically 
through the use of high accuracy direct numerical simulations
in order to advance our understanding of drop deformation and splashing dynamics in an aerodynamic context.
A model that encompasses the transition from larger lengthscales of typical engineering applications (of the order 
of an airfoil chord or nacelle diameter) down to the local impact region, whilst accounting 
for the surrounding (non-quiescent) air flow has been proposed. Using an oblique-stagnation 
point flow model for the background flow provides a suitable framework for the impinging 
drop to naturally interact with structures such as the growing boundary layers it would encounter 
on aircraft surfaces. Pre-impact deformation and break-up is compared to available experiments 
in the relevant parameter range of $\mathcal{O}(100)$ m/s impact velocities and $D=\mathcal{O}(100-1000)\ \mu$m diameter drops.
The impingement of the smallest droplets is described by a regular spreading motion, 
while beyond a certain size (estimated to be of approximately $50\ \mu$m), violent splashing is observed.
Topological changes such as droplet break-off or coalescence, as well as potential 
subsequent re-impingement of smaller fragments have been taken into account. 

Comparison with existing analytical results is possible for the tractable spreading dynamics of the smallest 
drops in the tested range (under $50\ \mu$m in diameter). Variation in the angle of impingement (not previously performed in this regime) 
reveals intriguing corner-type features at the advancing impact front which warrant further investigation.
These emerge as a result of the asymmetric impact pushing more fluid mass in a preferred direction, 
however with surface tension preventing break-up above a $30^{\circ}$ angle of incidence, and the action 
of the strong background flow induced shear. Below this angle,
fluid filaments have been shown to form and detach from the surface near the advancing droplet front, 
breaking into large droplets 
and being carried away by the surrounding air flow.
For larger drops, splashing and secondary drop ejection is captured numerically in detail 
and examined in order to advance understanding of the water retention process. 
The average number of secondary drops resulting from the impact increases with the initial droplet diameter,
with their sizes well-represented by a log-normal distribution, with a secondary local 
maximum emerging as a result of further break-up at the final stages of the simulated dynamics.

These findings provide detailed insight into the highly complex fluid dynamical 
processes occurring during aircraft flight through high liquid water content regions.
The present approach and simulations are a significant advance of
standard particle-based methods which are common industrial practice 
and which rely heavily on semi-empirical arguments. The modelled background air flow
provides a reliable local description of the flow in which a full interaction between 
the air and the liquid is permitted and the deformation of the drops is captured in detail before 
impact, while the emergence and movement of secondary drops in an active flow region 
is also treated realistically. Significant efforts have been made in order to ensure 
a highly accurate resolution of the flow. Nevertheless, we emphasise that these results 
originate from very intensive and resource heavy computational efforts (both in terms 
of runtime and data storage and processing requirements). Further advances in this area
of research in terms of both algorithms and raw computing power will facilitate more understanding 
of these violent regimes, with access to smaller grid sizes, larger domains (the issue of locality) 
as well as sensitivity to contact angle dynamics, being only a small subset 
of the possible future directions within this methodology.

In conclusion, this research provides a renewed perspective on the modelling of water 
catch on aircraft surfaces, with possible ramifications towards other areas involving 
high speed drop impact, such as inkjet printing, combustion and agricultural sprays. The presented results have shown very favourable agreement 
with recent experimental and analytical results, where possible, in an incredibly 
challenging regime, whilst new phenomena and detailed quantification of practical
information beyond the capabilities of present video technology and analytical treatments 
has also been provided. 
We believe that the proposed numerical framework is a valuable tool not only 
from the fundamental perspective in the study of drop impact, but also in an engineering context as 
a means of using scale transition to include detailed physical and fluid-related processes 
in water retention estimation and associated phenomena such as icing.

\section*{Acknowledgement}

The authors acknowledge the support of Innovate UK through the SANTANA (System Advances in Nacelle Technology AerodyNAmics, project reference 113001) program. 
The authors were also partly supported by EPSRC grants EP/K041134/1 and EP/L020564/1.
We are grateful to Richard Newman and Hui Yao from Bombardier Aerospace Ltd. for fruitful practical discussions.
Finally, we would like to thank the anonymous referees for their insight and thoughtful comments.

\section*{References}

\bibliography{Impact3D}

\providecommand{\noopsort}[1]{}\providecommand{\singleletter}[1]{#1}%
\begin{thebibliography}{70}
\expandafter\ifx\csname natexlab\endcsname\relax\def\natexlab#1{#1}\fi
\expandafter\ifx\csname url\endcsname\relax
  \def\url#1{\texttt{#1}}\fi
\expandafter\ifx\csname urlprefix\endcsname\relax\def\urlprefix{URL }\fi

\bibitem[{Afkhami et~al.(2009)Afkhami, Zaleski, and Bussman}]{afkhami1}
Afkhami, S., Zaleski, S., Bussman, M., 2009. A mesh-dependent model for
  applying dynamic contact angles to {VOF} simulations. J. Comput. Phys. 228,
  5370--5389.

\bibitem[{Agbaglah et~al.(2015)Agbaglah, Thoraval, Thoroddsen, Zhang, Fezzaa,
  and Deegan}]{Agbaglah1}
Agbaglah, G., Thoraval, M.-J., Thoroddsen, S., Zhang, L., Fezzaa, K., Deegan,
  R., 2 2015. Drop impact into a deep pool: vortex shedding and jet formation.
  J. Fluid Mech. 764.

\bibitem[{Bartolo et~al.(2005)Bartolo, Josserand, and
  Bonn}]{bartolo2005retraction}
Bartolo, D., Josserand, C., Bonn, D., 2005. Retraction dynamics of aqueous
  drops upon impact on non-wetting surfaces. Journal of Fluid Mechanics 545,
  329--338.

\bibitem[{Bilodeau et~al.(2015)Bilodeau, Habashi, Fossati, and
  Baruzzi}]{bilodeau2015}
Bilodeau, D., Habashi, W., Fossati, M., Baruzzi, G., 2015. Eulerian modeling of
  supercooled large droplet splashing and bouncing. Journal of Aircraft 52~(5),
  1611--1624.

\bibitem[{Bird et~al.(2009)Bird, Tsai, and Stone}]{bird2009inclined}
Bird, J.~C., Tsai, S.~S., Stone, H.~A., 2009. Inclined to splash: triggering
  and inhibiting a splash with tangential velocity. New Journal of Physics
  11~(6), 063017.

\bibitem[{Blyth and Pozrikidis(2005)}]{Blyth1}
Blyth, M., Pozrikidis, C., 2005. Stagnation-point flow against a liquid film on
  a plane wall. Acta Mechanica 180~(1-4), 203--219.

\bibitem[{Bragg(1996)}]{bragg1996aerodynamics}
Bragg, M., 1996. Aerodynamics of supercooled-large-droplet ice accretions and
  the effect on aircraft control. In: Proceedings of the FAA International
  Conference on Aircraft Inflight Icing. Vol.~2. pp. 387--399.

\bibitem[{Cheng and Lou(2015)}]{cheng2015numerical}
Cheng, M., Lou, J., 2015. A numerical study on splash of oblique drop impact on
  wet walls. Computers \& Fluids 115, 11--24.

\bibitem[{Cherdantsev et~al.(2017)Cherdantsev, Hann, Hewakandamby, and
  Azzopardi}]{Cherdantsev2017}
Cherdantsev, A.~V., Hann, D.~B., Hewakandamby, B.~N., Azzopardi, B.~J., 2017.
  Study of the impacts of droplets deposited from the gas core onto a
  gas-sheared liquid film. International Journal of Multiphase Flow 88, 69 --
  86.

\bibitem[{Clanet et~al.(2004)Clanet, B{\'e}guin, Richard, and
  Qu{\'e}r{\'e}}]{clanet2004maximal}
Clanet, C., B{\'e}guin, C., Richard, D., Qu{\'e}r{\'e}, D., 2004. Maximal
  deformation of an impacting drop. Journal of Fluid Mechanics 517, 199--208.

\bibitem[{Deng et~al.(2009)Deng, Varanasi, Hsu, Bhate, Keimel, Stein, and
  Blohm}]{Deng1}
Deng, T., Varanasi, K., Hsu, M., Bhate, N., Keimel, C., Stein, J., Blohm, M.,
  2009. Nonwetting of impinging droplets on textured surfaces. Applied Physics
  Letters 94~(13), 133109.

\bibitem[{Dorrepaal(1986)}]{Dorrepaal1}
Dorrepaal, J., 1986. An exact solution of the {N}avier-{S}tokes equation which
  describes non-orthogonal stagnation-point flow in two dimensions. J. Fluid
  Mech. 163, 141--147.

\bibitem[{Eggers et~al.(2010)Eggers, Fontelos, Josserand, and
  Zaleski}]{Eggers1}
Eggers, J., Fontelos, M., Josserand, C., Zaleski, S., 2010. Drop dynamics after
  impact on a solid wall: theory and simulations. Physics of Fluids
  (1994-present) 22~(6), 062101.

\bibitem[{Fedorchenko et~al.(2005)Fedorchenko, Wang, and
  Wang}]{fedorchenko2005effect}
Fedorchenko, A., Wang, A., Wang, Y., 2005. Effect of capillary and viscous
  forces on spreading of a liquid drop impinging on a solid surface. Physics of
  Fluids 17~(9), 093104.

\bibitem[{Fuster(2013)}]{fuster2013}
Fuster, D., 2013. An energy preserving formulation for the simulation of
  multiphase turbulent flows. J. Computational Phys. 235, 114--128.

\bibitem[{Gent et~al.(2000)Gent, Dart, and Cansdale}]{Gent1}
Gent, R., Dart, N., Cansdale, J., 2000. Aircraft icing. Phil. Trans. R. Soc. A
  358~(1776), 2873--2911.

\bibitem[{Guo et~al.(2016)Guo, Lian, and Sussman}]{guo2016investigation}
Guo, Y., Lian, Y., Sussman, M., 2016. Investigation of drop impact on dry and
  wet surfaces with consideration of surrounding air. Physics of Fluids 28~(7),
  073303.

\bibitem[{Hiemenz(1911)}]{Hiemenz1}
Hiemenz, K., 1911. Die {G}renzschicht an einem in den gleichf{\"o}rmigen
  {F}l{\"u}ssigkeitsstrom eingetauchten geraden {K}reiszylinder. Ph.D. thesis,
  Dinglers Polytech. J.

\bibitem[{Honsek et~al.(2008)Honsek, Habashi, and Aub{\'e}}]{honsek2008}
Honsek, R., Habashi, W., Aub{\'e}, M., 2008. Eulerian modeling of in-flight
  icing due to supercooled large droplets. Journal of Aircraft 45~(4),
  1290--1296.

\bibitem[{Howarth(1951)}]{howarth1951cxliv}
Howarth, L., 1951. The boundary layer in three dimensional flow.—part ii. the
  flow near a stagnation point. The {L}ondon, {E}dinburgh, and {D}ublin
  {P}hilosophical {M}agazine and {J}ournal of {S}cience 42~(335), 1433--1440.

\bibitem[{Jalaal and Mehravaran(2012)}]{Jalaal1}
Jalaal, M., Mehravaran, K., 2012. Fragmentation of falling liquid droplets in
  bag breakup mode. International Journal of Multiphase Flow 47~(0), 115 --
  132.

\bibitem[{Josserand and Thoroddsen(2016)}]{josserand2016drop}
Josserand, C., Thoroddsen, S., 2016. Drop impact on a solid surface. Annu. Rev.
  Fluid Mech. 48, 365--391.

\bibitem[{Jung and Hutchings(2012)}]{jung2012}
Jung, S., Hutchings, I., 2012. The impact and spreading of a small liquid drop
  on a non-porous substrate over an extended time scale. Soft Matter 8~(9),
  2686--2696.

\bibitem[{Kim(2007)}]{kim2007}
Kim, J., 2007. Spray cooling heat transfer: the state of the art. Int. J. Heat
  and Fluid Flow 28~(4), 753--767.

\bibitem[{Mandre and Brenner(2012)}]{mandre2012}
Mandre, S., Brenner, M., 2012. The mechanism of a splash on a dry solid
  surface. J. Fluid Mech. 690, 148--172.

\bibitem[{Marston et~al.(2012)Marston, Zhu, Vakarelski, and
  Thoroddsen}]{marston2012}
Marston, J., Zhu, Y., Vakarelski, I., Thoroddsen, S., 2012. Deformed liquid
  marbles: {F}reezing drop oscillations with powders. Powder Tech. 228,
  424--428.

\bibitem[{Ming and Jing(2014)}]{ming2014lattice}
Ming, C., Jing, L., 2014. Lattice {B}oltzmann simulation of a drop impact on a
  moving wall with a liquid film. Computers \& Mathematics with Applications
  67~(2), 307--317.

\bibitem[{Moreira et~al.(2010)Moreira, Moita, and Panao}]{Moreira1}
Moreira, A., Moita, A., Panao, M., 2010. Advances and challenges in explaining
  fuel spray impingement: How much of single droplet impact research is useful?
  Progress in energy and combustion science 36~(5), 554--580.

\bibitem[{Mundo et~al.(1995)Mundo, Sommerfeld, and Tropea}]{Mundo1}
Mundo, C., Sommerfeld, M., Tropea, C., 1995. Droplet-wall collisions:
  Experimental studies of the deformation and breakup process. Int. J.
  Multiphase Flow 21~(2), 151 -- 173.

\bibitem[{Papadakis et~al.(2003)Papadakis, Rachman, Wong, Bidwell, and
  Bencic}]{papadakis2003}
Papadakis, M., Rachman, A., Wong, S.-C., Bidwell, C., Bencic, T., 2003. An
  experimental investigation of {SLD} impingement on airfoils and simulated ice
  shapes. Tech. rep., SAE Technical Paper.

\bibitem[{Papadakis et~al.(2004)Papadakis, Rachman, Wong, Yeong, Hung, and
  Bidwell}]{papadakisExp}
Papadakis, M., Rachman, A., Wong, S.-C., Yeong, H.-W., Hung, K., Bidwell, C.,
  2004. Water impingement experiments on a {NACA} 23012 airfoil with simulated
  glaze ice shapes. AIAA Paper 565, 1--40.

\bibitem[{Pasandideh-Fard et~al.(1996)Pasandideh-Fard, Qiao, Chandra, and
  Mostaghimi}]{Fard1}
Pasandideh-Fard, M., Qiao, Y., Chandra, S., Mostaghimi, J., 1996. Capillary
  effects during droplet impact on a solid surface. Physics of Fluids
  (1994-present) 8~(3), 650--659.

\bibitem[{Philippi et~al.(2016)Philippi, Lagr{\'e}e, and
  Antkowiak}]{philippi2016drop}
Philippi, J., Lagr{\'e}e, P.-Y., Antkowiak, A., 2016. Drop impact on a solid
  surface: short-time self-similarity. Journal of Fluid Mechanics 795, 96--135.

\bibitem[{Popinet(2003)}]{popinet1}
Popinet, S., 2003. Gerris: A tree-based adaptive solver for the incompressible
  {E}uler equations in complex geometries. J. Comput. Phys. 190, 572.

\bibitem[{Popinet(2009)}]{popinet2}
Popinet, S., 2009. An accurate adaptive solver for surface-tension-driven
  interfacial flows. J. Comput. Phys. 228, 5838.

\bibitem[{Potapczuk et~al.(1993)Potapczuk, Al-Khalil, and
  Velazquez}]{potapczuk1993ice}
Potapczuk, M., Al-Khalil, K., Velazquez, M., 1993. Ice accretion and
  performance degradation calculations with lewice/ns. In: $31$st Aerospace
  Sciences Meeting. p. 173.

\bibitem[{Rein(1993)}]{rein1993}
Rein, M., 1993. Phenomena of liquid drop impact on solid and liquid surfaces.
  Fluid Dyn. Res. 12~(2), 61.

\bibitem[{Riboux and Gordillo(2014)}]{riboux2014experiments}
Riboux, G., Gordillo, J., 2014. Experiments of drops impacting a smooth solid
  surface: a model of the critical impact speed for drop splashing. Physical
  review letters 113~(2), 024507.

\bibitem[{Riboux and Gordillo(2015)}]{riboux2015diameters}
Riboux, G., Gordillo, J., 2015. The diameters and velocities of the droplets
  ejected after splashing. Journal of Fluid Mechanics 772, 630--648.

\bibitem[{Roisman(2009)}]{Roisman2009}
Roisman, I., 2009. Inertia dominated drop collisions. ii. an analytical
  solution of the {N}avier–{S}tokes equations for a spreading viscous film.
  Physics of Fluids 21~(5), 052104.

\bibitem[{Rutkowski et~al.(2003)Rutkowski, Wright, and
  Potapczuk}]{rutkowski2003numerical}
Rutkowski, A., Wright, W., Potapczuk, M., 2003. Numerical study of droplet
  splashing and re-impingement. In: 41st Aerospace Sciences Meeting and
  Exhibit. p. 388.

\bibitem[{Sawan and Carbon(1975)}]{sawan1975}
Sawan, M., Carbon, M., 1975. A review of spray-cooling and bottom-flooding work
  for {LWR} cores. Nuclear Eng. Design 32~(2), 191--207.

\bibitem[{Schroll et~al.(2010)Schroll, Josserand, Zaleski, and
  Zhang}]{Schroll1}
Schroll, R., Josserand, C., Zaleski, S., Zhang, W., Jan 2010. Impact of a
  {V}iscous {L}iquid {D}rop. Phys. Rev. Lett. 104, 034504.

\bibitem[{Sikalo et~al.(2005)Sikalo, Tropea, and Ganic}]{Sikalo1}
Sikalo, S., Tropea, C., Ganic, E., 2005. Impact of droplets onto inclined
  surfaces. Journal of Colloid and Interface Science 286~(2), 661 -- 669.

\bibitem[{Sor and Garc{\'\i}a-Magari{\~n}o(2015)}]{sor2015modeling}
Sor, S., Garc{\'\i}a-Magari{\~n}o, A., 2015. Modeling of droplet deformation
  near the leading edge of an airfoil. Journal of Aircraft 52~(6), 1838--1846.

\bibitem[{Stow and Hadfield(1981)}]{stow1981}
Stow, C., Hadfield, M., 1981. An experimental investigation of fluid flow
  resulting from the impact of a water drop with an unyielding dry surface.
  Proc. of the Royal Soc. of London. Series A 373~(1755), 419--441.

\bibitem[{Stuart(1959)}]{Stuart1}
Stuart, J., 1959. The viscous flow near a stagnation point when the external
  flow has uniform vorticity. J. Aerosp. Sci. 26~(124).

\bibitem[{Tamada(1979)}]{Tamada1}
Tamada, K., 1979. Two-dimensional stagnation-point flow impinging obliquely on
  a plane wall. J. Phys. Soc. of Japan 46, 310.

\bibitem[{Thoraval(2013)}]{thoravalThesis}
Thoraval, M.-J., March 2013. Drop impact splashing and air entrapment. Ph.D.
  thesis, KAUST.

\bibitem[{Thoraval et~al.(2012)Thoraval, Takehara, Etoh, Popinet, Ray,
  Josserand, Zaleski, and Thoroddsen}]{Thoraval2}
Thoraval, M.-J., Takehara, K., Etoh, T., Popinet, S., Ray, P., Josserand, C.,
  Zaleski, S., Thoroddsen, S., 2012. von {K}{\'a}rm{\'a}n vortex street within
  an impacting drop. Phys. Rev. Lett. 108~(26), 264506.

\bibitem[{Thoroddsen et~al.(2008)Thoroddsen, Etoh, and
  Takehara}]{thoroddsen2008}
Thoroddsen, S., Etoh, T., Takehara, K., 2008. High-speed imaging of drops and
  bubbles. Annu. Rev. Fluid Mech. 40, 257--285.

\bibitem[{Thoroddsen and Shen(2001)}]{thoroddsen2001}
Thoroddsen, S., Shen, A., 2001. Granular jets. Physics of Fluids (1994-present)
  13~(1), 4--6.

\bibitem[{Tilley and Weidman(1998)}]{tilley1998oblique}
Tilley, B., Weidman, P., 1998. Oblique two-fluid stagnation-point flow.
  European Journal of Mechanics-B/Fluids 17~(2), 205--217.

\bibitem[{Tooke and Blyth(2008)}]{Tooke1}
Tooke, R., Blyth, M., 2008. A note on oblique stagnation-point flow. Physics of
  Fluids (1994-present) 20~(3).

\bibitem[{Tryggvason et~al.(2011)Tryggvason, Scardovelli, and
  Zaleski}]{tryggvason1}
Tryggvason, G., Scardovelli, R., Zaleski, S., 2011. Direct numerical
  simulations of gas-liquid multiphase flows. Cambridge University Press.

\bibitem[{Tsai et~al.(2009)Tsai, Pacheco, Pirat, Lefferts, and Lohse}]{Tsai2}
Tsai, P., Pacheco, S., Pirat, C., Lefferts, L., Lohse, D., 2009. Drop impact
  upon micro- and nanostructured superhydrophobic surfaces. Langmuir 25~(20),
  12293--12298.

\bibitem[{van Dam and Le~Clerc(2004)}]{vanDam1}
van Dam, D., Le~Clerc, C., 2004. Experimental study of the impact of an ink-jet
  printed droplet on a solid substrate. Physics of Fluids (1994-present)
  16~(9), 3403--3414.

\bibitem[{Vargas et~al.(2012)Vargas, Sor, and
  Garc{\'\i}a~Magari{\~n}o}]{vargas2012mechanism}
Vargas, M., Sor, S., Garc{\'\i}a~Magari{\~n}o, A., 2012. Mechanism of water
  droplet breakup near the leading edge of an airfoil. In: 4th AIAA Atmospheric
  and Space Environments Conference. p. 3129.

\bibitem[{Visser et~al.(2015)Visser, Frommhold, Wildeman, Mettin, Lohse, and
  Sun}]{Visser1}
Visser, C., Frommhold, P., Wildeman, S., Mettin, R., Lohse, D., Sun, C., 2015.
  Dynamics of high-speed micro-drop impact: numerical simulations and
  experiments at frame-to-frame times below 100 ns. Soft Matter 11, 1708--1722.

\bibitem[{Wang(2008)}]{wang2008similarity}
Wang, C., 2008. Similarity stagnation point solutions of the {N}avier--{S}tokes
  equations--review and extension. European Journal of Mechanics-B/Fluids
  27~(6), 678--683.

\bibitem[{Wildeman et~al.(2016)Wildeman, Visser, Sun, and
  Lohse}]{wildeman2016spreading}
Wildeman, S., Visser, C., Sun, C., Lohse, D., 2016. On the spreading of
  impacting drops. Journal of Fluid Mechanics 805, 636--655.

\bibitem[{Worthington(1876)}]{worthington1876forms}
Worthington, A., 1876. On the forms assumed by drops of liquids falling
  vertically on a horizontal plate. Proceedings of the {R}oyal {S}ociety of
  {L}ondon 25~(171-178), 261--272.

\bibitem[{Worthington(1908)}]{worthington1908}
Worthington, A., 1908. A study of splashes. Longmans, Green, and Co.

\bibitem[{Wright(2005)}]{wright2005validation}
Wright, W., 2005. Validation results for lewice 3.0. In: $43$rd AIAA Aerospace
  Sciences Meeting and Exhibit. p. 1243.

\bibitem[{Wright(2006)}]{wright2006}
Wright, W., 2006. Further refinement of the lewice sld model. In: 44th AIAA
  Aerospace Sciences Meeting and Exhibit. p. 464.

\bibitem[{Wright and Potapczuk(2004)}]{wright2004}
Wright, W., Potapczuk, M., 2004. Semi-empirical modelling of sld physics. In:
  42nd AIAA Aerospace Sciences Meeting and Exhibit. p. 412.

\bibitem[{Xie et~al.(2017)Xie, Hewitt, Pavlidis, Salinas, Pain, and
  Matar}]{Xie2017}
Xie, Z., Hewitt, G.~F., Pavlidis, D., Salinas, P., Pain, C.~C., Matar, O.~K.,
  2017. Numerical study of three-dimensional droplet impact on a flowing liquid
  film in annular two-phase flow. Chemical Engineering Science 166, 303--312.

\bibitem[{Xu et~al.(2005)Xu, Zhang, and Nagel}]{Xu1}
Xu, L., Zhang, W., Nagel, S., 2005. Drop splashing on a dry smooth surface.
  Phys. Rev. Lett. 94, 184505.

\bibitem[{Yarin(2006)}]{Yarin1}
Yarin, A., 2006. Drop impact dynamics: splashing, spreading, receding,
  bouncing... Annu. Rev. Fluid Mech. 38, 159--192.

\bibitem[{Yokoi et~al.(2009)Yokoi, Vadillo, Hinch, and
  Hutchings}]{yokoi2009numerical}
Yokoi, K., Vadillo, D., Hinch, J., Hutchings, I., 2009. Numerical studies of
  the influence of the dynamic contact angle on a droplet impacting on a dry
  surface. Physics of Fluids 21~(7), 072102.

\end{thebibliography}

\end{document}